\documentclass[12pt,twoside]{article}

\usepackage{url}
\usepackage[pdftex]{graphicx}
\DeclareGraphicsExtensions{.pdf} 
\usepackage{amssymb}
\usepackage[pdftex,                
bookmarks=true,
bookmarksnumbered=true,
hypertexnames=false,
breaklinks=true,
linkbordercolor={0 0 1}]{hyperref} 
\hypersetup{
pdftitle    = {Automated identification of neurons and their locations},
pdfsubject  = {journal article},
pdfkeywords = {neuron identification, active contours, artificial neural network, training, Nissl-stain},
}
\usepackage[colon,authoryear,round]{natbib}

\begin{document}


\title{Automated identification of neurons and their locations}

\maketitle

\noindent{
A.~Inglis$^{1 \ast}$, L.~Cruz$^1$,
D.~L.~Roe$^2$, H.~E.~Stanley$^1$, D.~L.~Rosene$^{2,3}$, and B.~Urbanc$^1$} \medskip

\noindent
{\scriptsize
$^1$Center for Polymer Studies, Department of Physics, Boston
University, Boston, MA 02215 \\
$^2$Department of Anatomy and Neurobiology, Boston University School of Medicine, Boston, MA 02118 \\
$^3$Yerkes National Primate Research Center, Emory University, Atlanta GA, 30322 \\
$^\ast$Corresponding author: ainglis@physics.bu.edu }


\begin{abstract}
Individual locations of many neuronal cell bodies ($>10^4$) are needed to
enable statistically significant measurements of spatial organization within
the brain such as nearest-neighbor and microcolumnarity measurements. In this
paper, we introduce an Automated Neuron Recognition Algorithm (ANRA) which
obtains the (x,y) location of individual neurons within digitized images of
Nissl-stained, 30 micron thick, frozen sections of the cerebral cortex of the
Rhesus monkey. Identification of neurons within such Nissl-stained sections
is inherently difficult due to the variability in neuron staining, the
overlap of neurons, the presence of partial or damaged neurons at tissue
surfaces, and the presence of non-neuron objects, such as glial cells, blood
vessels, and random artifacts. To overcome these challenges and identify
neurons, ANRA applies a combination of image segmentation and machine
learning. The steps involve active contour segmentation to find outlines of
potential neuron cell bodies followed by artificial neural network training
using the segmentation properties (size, optical density, gyration, etc.) to
distinguish between neuron and non-neuron segmentations. ANRA positively
identifies $86\pm5\%$ neurons with $15\pm8\%$ error (mean $\pm$ st.dev.) on a
wide range of Nissl-stained images, whereas semi-automatic methods obtain
$80\pm7\%/17\pm12\%$. A further advantage of ANRA is that it affords an
unlimited increase in speed from semi-automatic methods, and is
computationally efficient, with the ability to recognize $\sim$100 neurons
per minute using a standard personal computer. ANRA is amenable to analysis
of huge photo-montages of Nissl-stained tissue, thereby opening the door to
fast, efficient and quantitative analysis of vast stores of archival material
that exist in laboratories and research collections around the world. The ANRA software is available at \url{http://physics.bu.edu/~ainglis/ANRA/}.

\end{abstract}

\section{\label{sec:introduction}Introduction}

Since the 1980s, the application of unbiased stereological approaches to
quantify objects of biological interest has allowed for rigorous measurements
of many parameters of brain structure including total neuron number, area,
and volume.  These approaches are based on systematic random sampling from
defined regions of interest using unbiased
estimators~\citep{Mayhew:1991,Hof:2004}.  While these measurements have
produced extremely valuable insights into the structural organization of the
brain, including age-related preservation of neuron
numbers~\citep{Peters:1998}, these ``first order'' stereological parameters
only partially describe the structural organization of the brain, as they
cannot efficiently quantify ``second order'' parameters that measure more
complex spatial properties of neuron organization, such as the nearest
neighbor arrangement
\citep{Enoch:1996,Schmitz:2002,Duyckaerts:2000,Krasnoperov:2004,Hof:2003,Urbanc:2002}
and arrangement into mini- or microcolumns
\citep{Cruz:2005,Buldyrev:2000,Buxhoeveden:1996}.

Several approaches can be used to quantify ``second order''
parameters. Stereological methods can quantify nearest-neighbor arrangement
\citep{Schmitz:2002}, but the methods are labor intensive and would be
difficult to apply to large brain areas.  Image Fourier methods do not
require manual marking of neuron locations and can quantify ``vertical bias''
of objects within an image~\citep{Casanova:2006}, but do not discern between
the contribution from glial and neuronal cell bodies.

Alternatively, pair correlation methods use concepts from statistical physics
to calculate correlation properties such as 1D nearest-neighbor
\citep{Urbanc:2002} and 2D microcolumnar organization~\citep{Buldyrev:2000}
of neurons, as well as more discerning properties of spatial arrangement,
such as the strength of microcolumnar order and microcolumnar width and
length~\citep{Cruz:2005}. The multitude of spatial organization quantities
that can be calculated with pair correlation analysis makes it appealing to
apply to large brain areas.  To do that, we first need to address the major
challenge to this approach: how to obtain the necessarily large number of
neuron locations ($10^3-10^4$ locations per measurement) to get statistically
significant results (see Sec.~\ref{sec:dmm} and Discussion) over large
regions of the brain, reaching $\sim10^6$ for a large study.  The acquisition
of such numbers of neurons by manually or semi-automatically identifying and
marking the location of each is prohibitively time-consuming and open to user
bias.  Hence, correlative analysis of spatial relationships among neurons (as
well as non-stereology based cell counts~\citep{Benes:2005}) would be
dramatically facilitated by an automatic method for identifying and locating
the visible centers of neurons accurately and efficiently.

While various other immunhistochemical methods could facilitate automated
discrimination of neurons and glia better than Nissl, there are important
advantages to develop automated methods for Nissl-stained
tissue. Nissl-staining is the least expensive, easiest applied method for
staining both neurons and glia. Furthermore, there are thousands of unique
and often irreproducible collections of Nissl-stained brain material in
clinical and research labs around the world that could be analyzed using the
ANRA.

There are several challenges to automatically retrieve neuron locations from
two-dimensional digitized images of Nissl-stained brain tissue
(Fig.~\ref{issue_pic}{\bf a}).  A major challenge is to distinguish between
neuron and non-neuron objects, including staining errors, tissue folds, and
dirt particles, as well as blood vessels and glial cells.  Another challenge
is to identify neurons that differ almost as widely from each other as they
do from non-neuronal objects.  Neuron cell bodies are naturally diverse in
size and shape and have different orientations with respect to their dendrite
and axon processes.  Neurons can also be cleaved at the cutting surface or
damaged by the cutting process, which affects their shape in the tissue.
These variables lead to diverse neuron cell profiles within the tissue slice.
A further challenge is to discriminate between neurons that overlap, a common
finding as tissue sections are 3D volumes projected onto a
2D image.

There are currently several published approaches to automatic retrieval of
cell bodies from images.  Some methods use segmentation techniques based on
thresholding~\citep{Slater:1996, Benali:2003}, Potts model~\citep{Peng:2003},
watershed~\citep{Lin:2005}, and active contours~\citep{Ray:2002}.  Others use
trained neural networks to mark appropriately sized ``pixel patches'' as
cells of interest.  The ``pixel patch'' training methods use artificial
neural networks~\citep{Sjostrom:1999}, local linear mapping
\citep{Nattkemperer:2001}, Fischer's linear discriminant~\citep{Long:2005},
and support vector machines~\citep{Long:2006}.  Another method based on
template matching has been recently introduced by \citet{Costa:2006}.

\begin{figure}[p]
\begin{center}
$\begin{array}{c}
\includegraphics*[width=11cm]{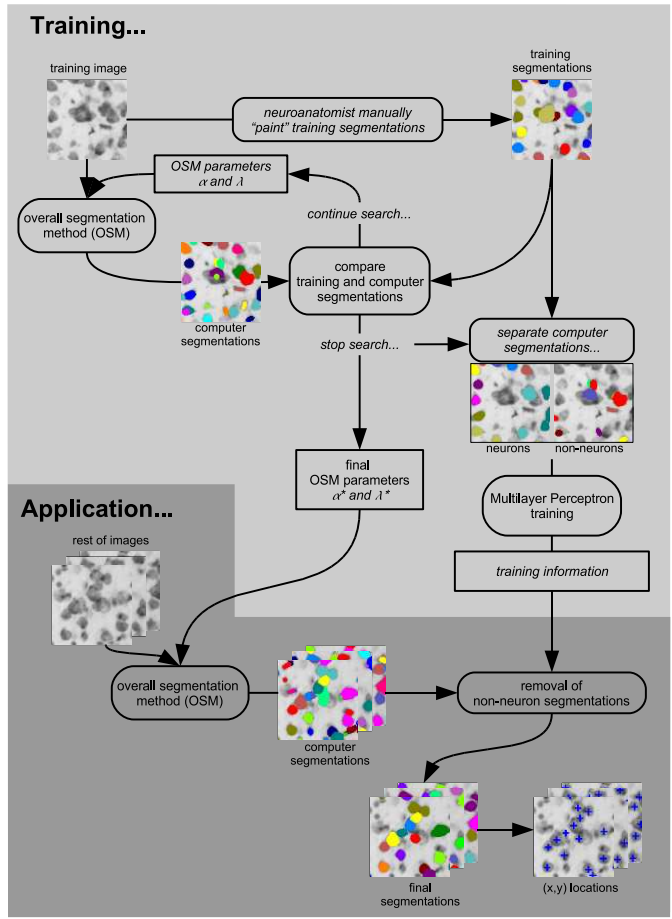}
\\
\end{array}$
\end{center}
\caption{A schematic diagram showing processes involved in the Automated
Neuron Recognition Algorithm (ANRA).  The schematic describes the two main
steps of the algorithm: training and application. Rectangles denote parameters
that pass through the algorithm. Ovals, such as the OSM, are the
computational parts of the algorithm, which can have images, segmentations, and parameters as their inputs
and outputs.}
\label{flow}
\end{figure}

In this paper we introduce and test an Automatic Neuron Recognition Algorithm
(ANRA) (Fig.~\ref{flow}) which uses a combination of segmenting and training to overcome the
challenges of retrieving neuron location in Nissl-stained tissue sections.
ANRA automatically identifies neurons from digital images and retrieves their
(x,y) locations.  
\bigskip

\section{\label{sec:methods}Methods}

\subsection{\label{sec:methods:-1}Image Input and Preprocessing}

\begin{figure}[p]
\begin{center}
$\begin{array}{c c}
\mbox{({\bf a})} & \begin{array}{c} \includegraphics*[width=8cm]{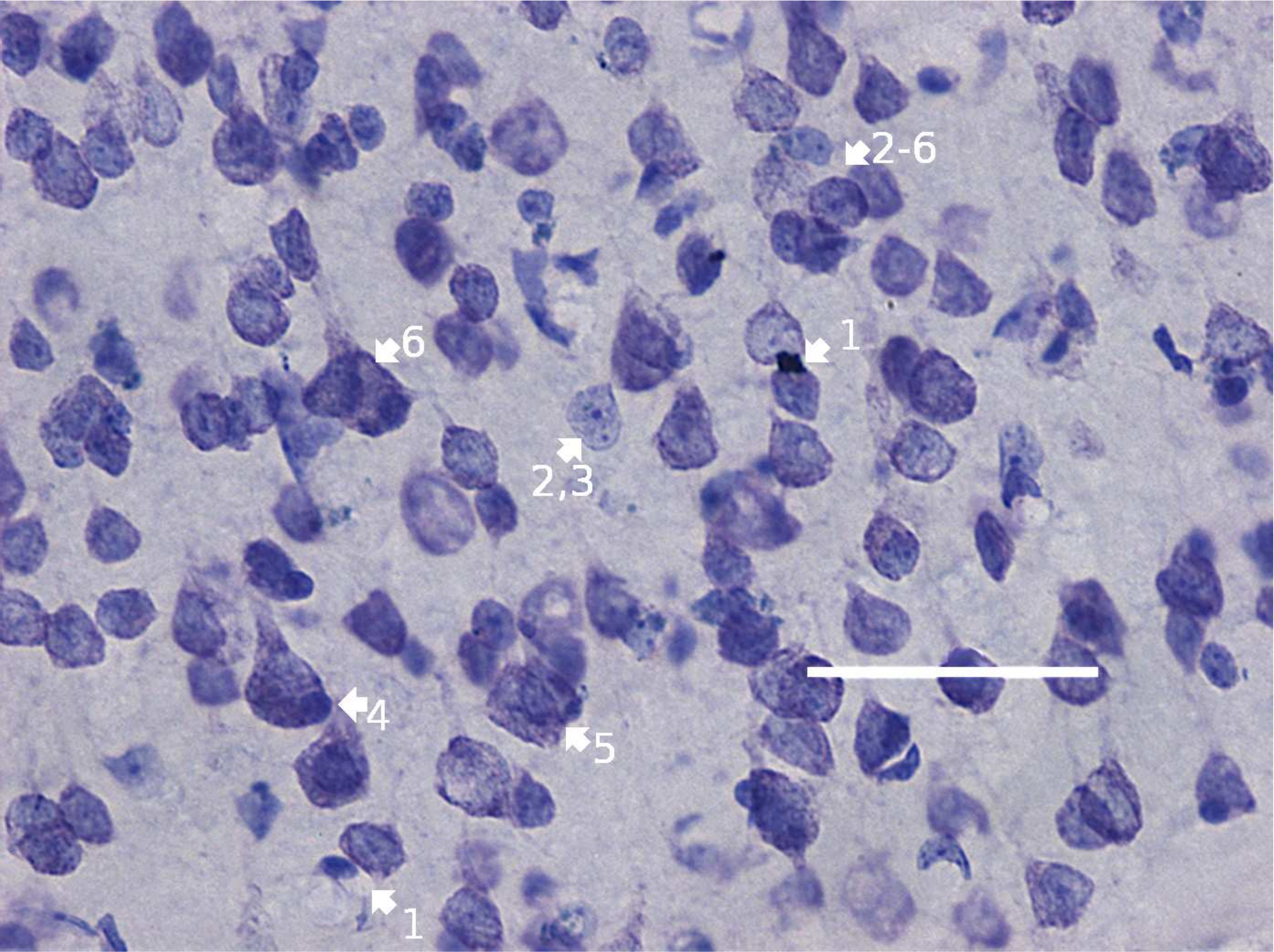} \end{array} \\
\mbox{({\bf b})} & \begin{array}{c} \includegraphics*[width=8cm]{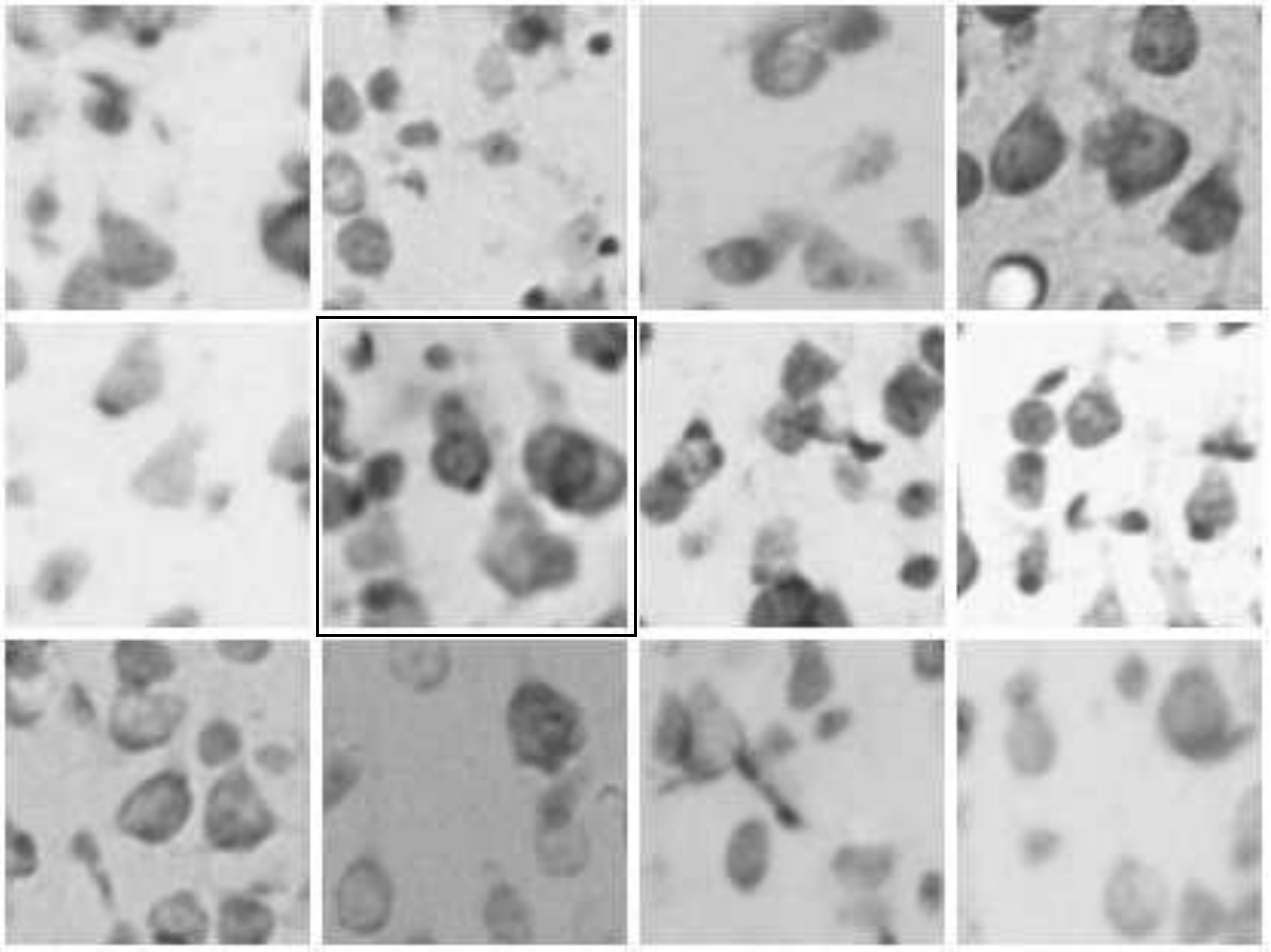} \end{array}
\end{array}$
\end{center}
\caption{Challenges of automated neuron recognition. ({\bf a}) 20x micrograph
(scale bar: $50{\mu}m$) of a typical section showing the difficulties of
separating neurons from glial cells and other artifacts in Nissl-stained
tissue: 1. capillaries, and unidentified material, 2. large glia
(astrocytes), 3. glial as light as neurons in some cases, 4. neurons
overlapped by glia (oligodendrocytes), 5. neurons overlapped by other
neurons, 6. multiple neurons and glial overlapped. ({\bf b}) 10x micrograph
examples showing varying image quality. The highlighted micrograph is
selected as an ``ideal'' contrast to be used in image normalization.}
\label{issue_pic}
\end{figure}

The inputs for ANRA are photomicrographs of 30 micron thick Nissl-stained
tissue section taken at 10x magnification and a resolution of 1.5 microns per
pixel. Because the 30 micron tissue section shrinks during processing to a
thickness of less than 10 microns, all of the tissue is in focus when viewed
at microscopic magnifications of 20X or lower, thus the 2D image properly
represents neuron locations . Since the color information is not as useful in
the monotone Nissl-stained images (Fig.~\ref{issue_pic}{\bf a}) the images
are converted to gray scale images ranging from 0~(black) to 255~(white).

\begin{figure}[p]
\begin{center}
$\begin{array}{c c}
\mbox{({\bf a})} & \begin{array}{c} \includegraphics*[width=5cm]{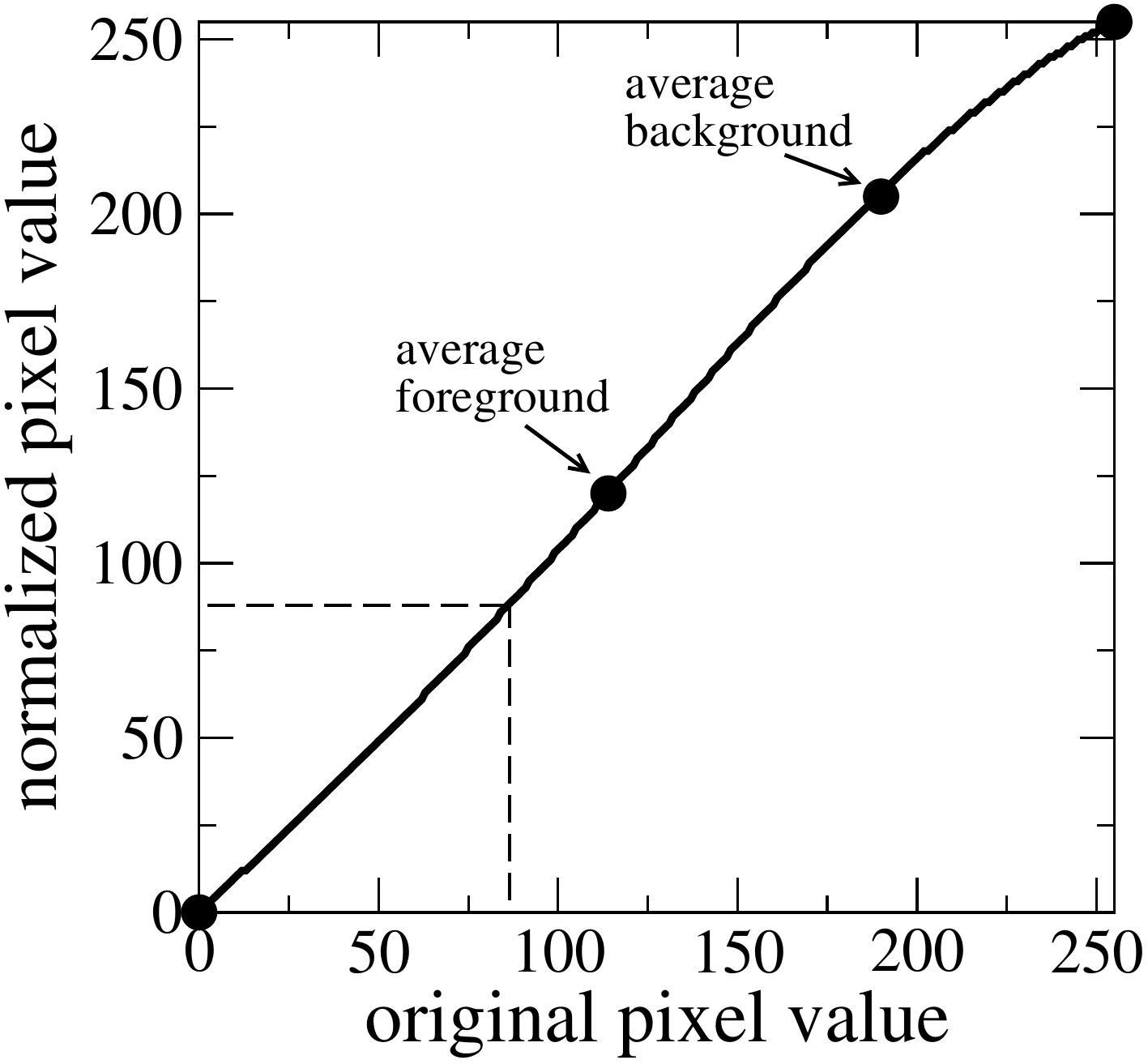} \end{array} \\
\mbox{({\bf b})} & \begin{array}{c} \includegraphics*[width=8cm]{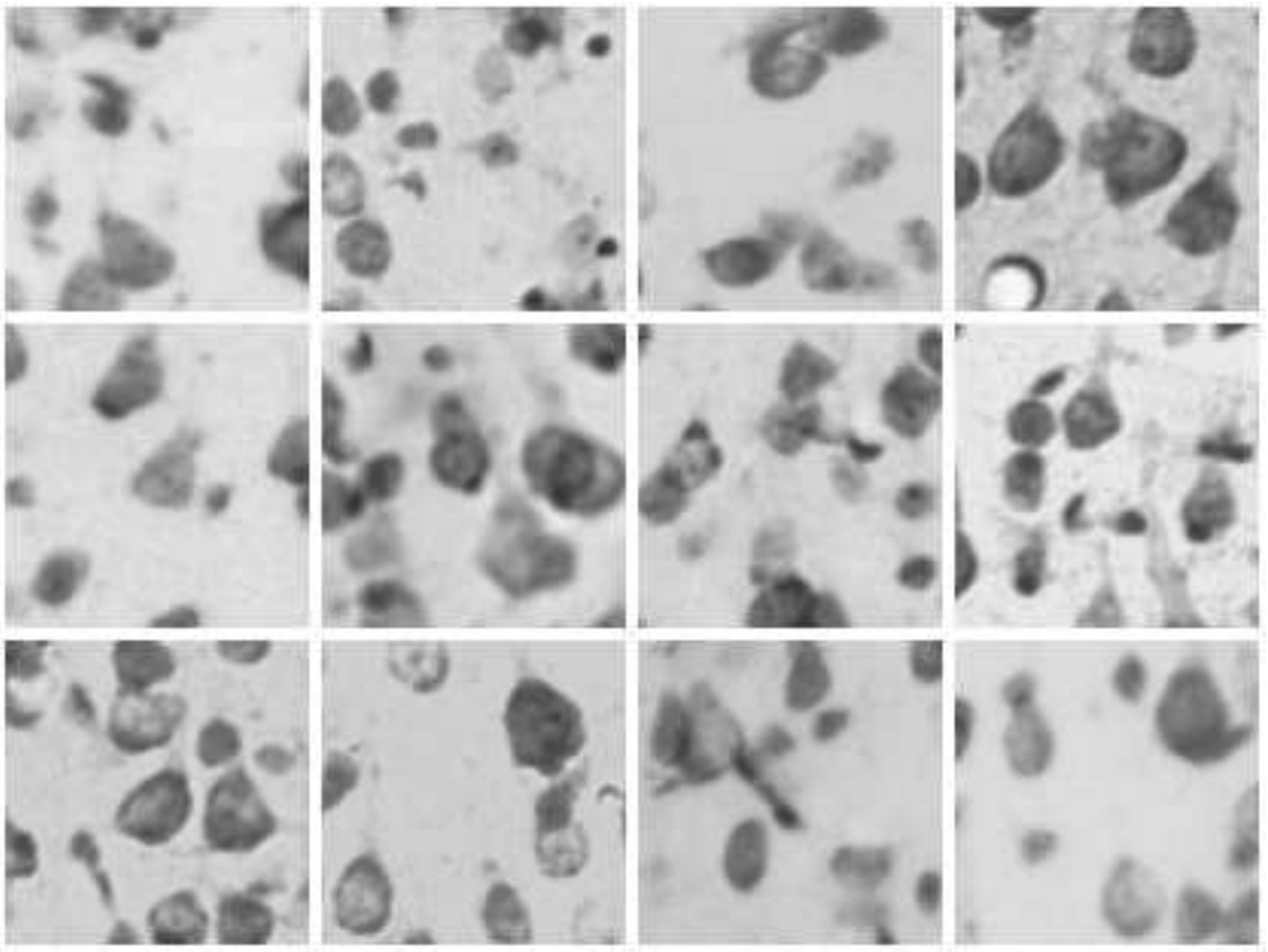} \end{array}
\end{array}$
\end{center}
\caption{({\bf a}) Preprocessing ``normalizes'' the images so that they every
image has the same background and foreground average optical densities,
thereby removing the challenge of varying image type within Nissl-stained
tissue. This is done by mapping optical density values of non-ideal images to
an ideal image so that the average foreground and background averages are the
same. The graph shows the optical density ranges of the ideal and non-ideal
images (0..255), and a Bezier curve that passes through 4 points: (0,0), the
background and foreground averages of the ideal and non-ideal images, and
(255,255). ({\bf b}) Examples of image normalization.}
\label{normalize}
\end{figure}

The photomicrographs are taken from different areas of the brain from
different subjects at different times. Therefore, images are of different
``quality'', reflecting a combination of variations in morphology, staining,
slide preparation, and digitization (Fig.~\ref{issue_pic}{\bf b}). To reduce
this variability, the images are first ``normalized'' such that every image
has the same background and foreground average optical density. This is done
by thresholding each image into foreground and background pixels and finding
the average optical density for the foreground and background separately. For
each image, the optical density histogram is then shifted to match the
foreground/background averages of an ideal image (Fig.~\ref{normalize}{\bf
a}). Fig.~\ref{normalize}{\bf b} shows the images final normalization as
compared to the original images in Fig.~\ref{issue_pic}{\bf b}.  This
preprocessing step removes most of the image variations due to processing
(staining, slide preparation, digitization, etc.) and is a key step toward
applying ANRA to an unlimited number of images that do not vary drastically
in intrinsic morphological differences (neuron density, shape, size,
etc.). There is no need for other preprocessing steps such as blurring or
sharpening since ANRA, by design, overcomes the challenges of noisy images
and weak boundary information.

\subsection{\label{sec:methods:0}Main segmentation tool: OSM}

Here we describe the segmentation procedure presented in Fig.~\ref{flow},
called the overall segmentation method (OSM).

\subsubsection{\label{0a}Over-marking the image}

\begin{figure}[p]
\begin{center}
$\begin{array}{c c}
\mbox{({\bf a})} & \begin{array}{c} \includegraphics*[width=3cm]{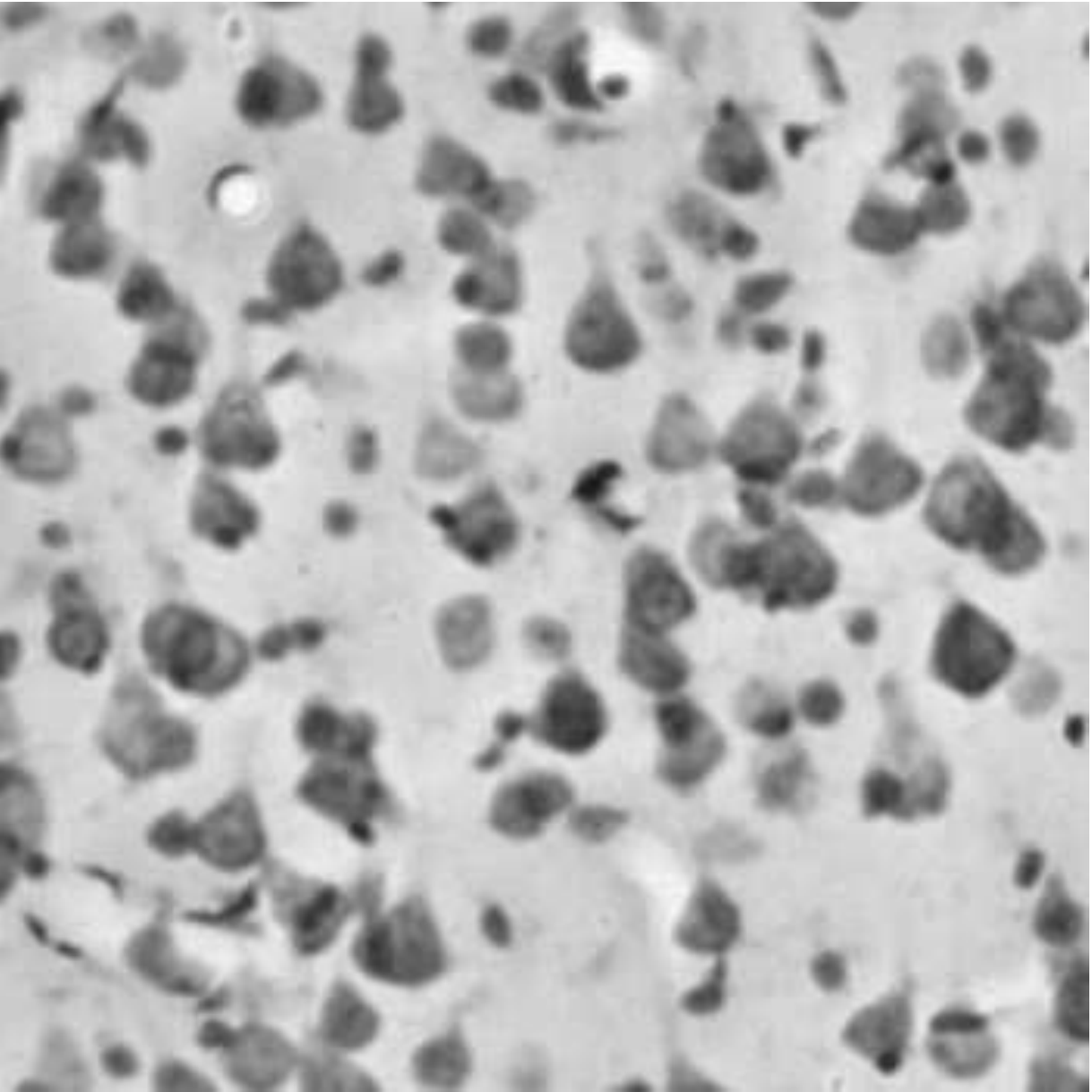}~\Rightarrow~\includegraphics*[width=3cm]{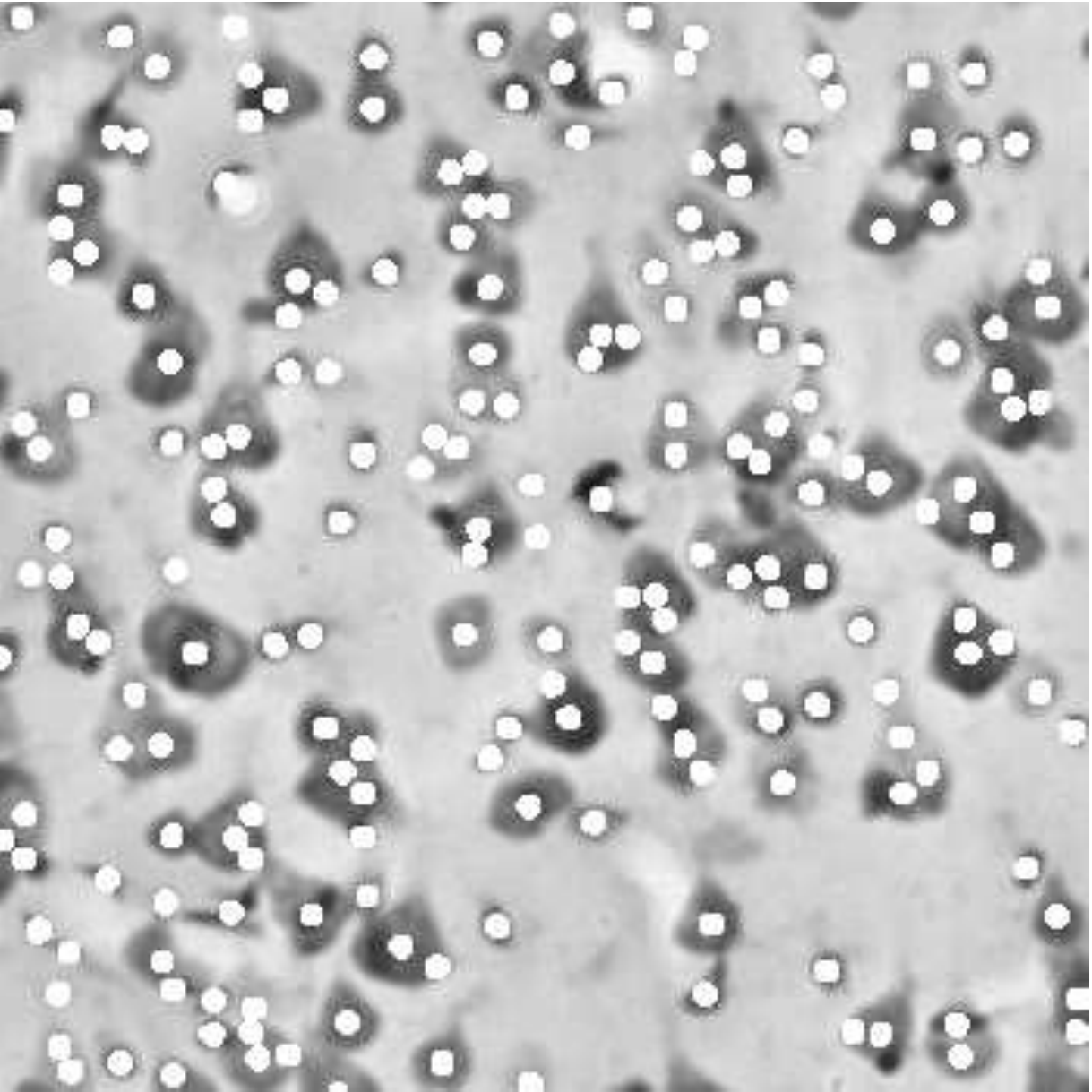} \end{array}\\
\mbox{({\bf b})} & \begin{array}{c} \includegraphics*[width=2cm]{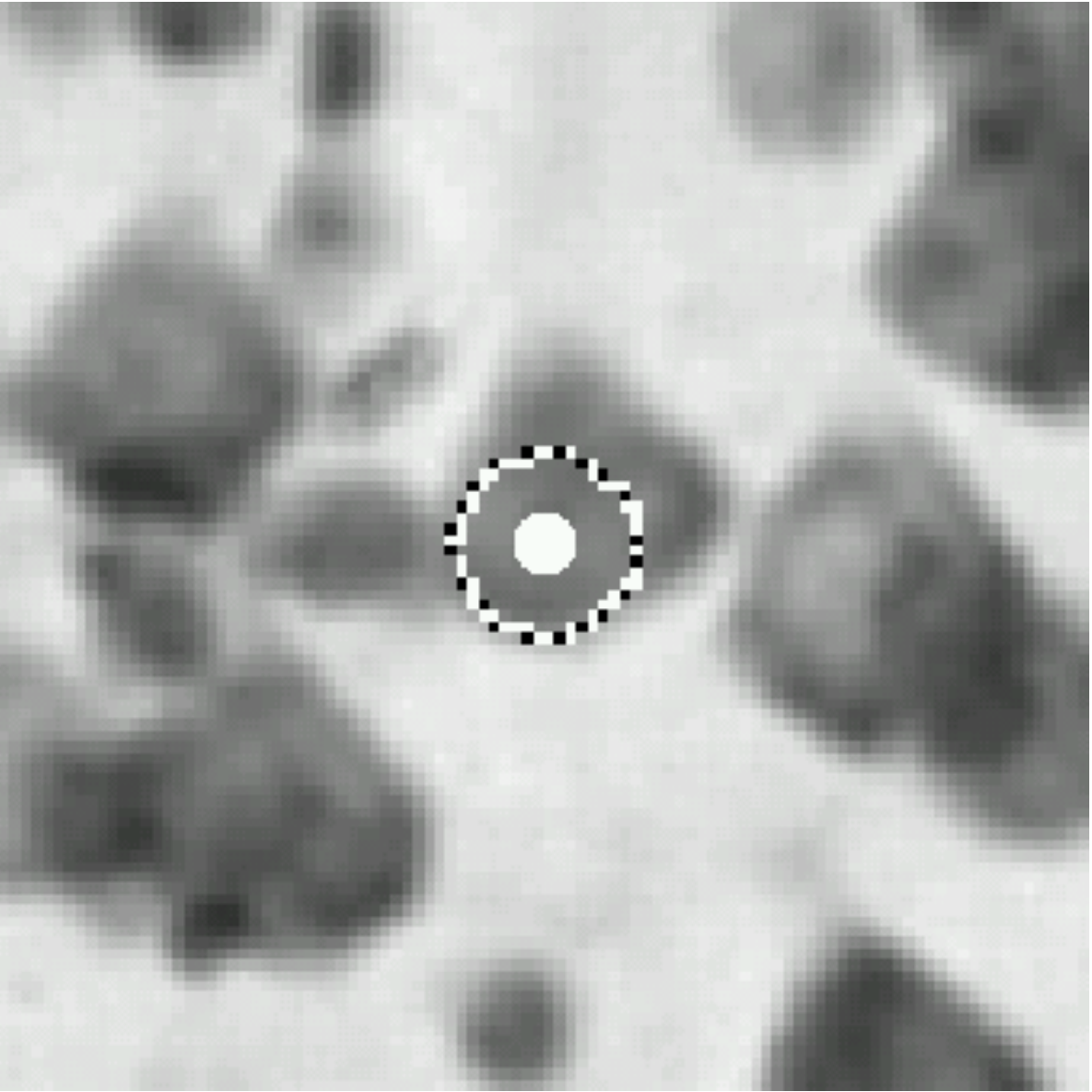}~\Rightarrow~\includegraphics*[width=2cm]{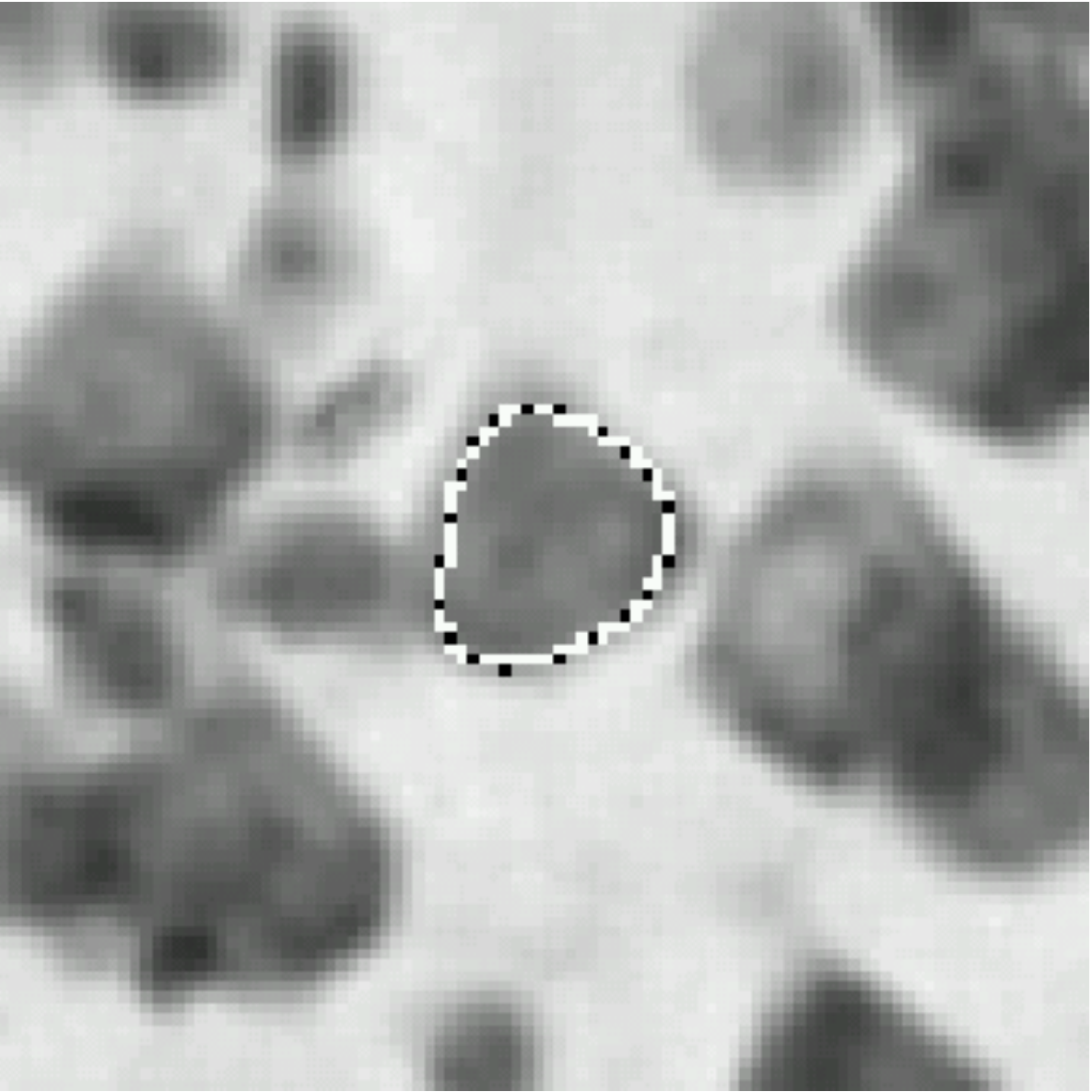}~\Rightarrow~\includegraphics*[width=2cm]{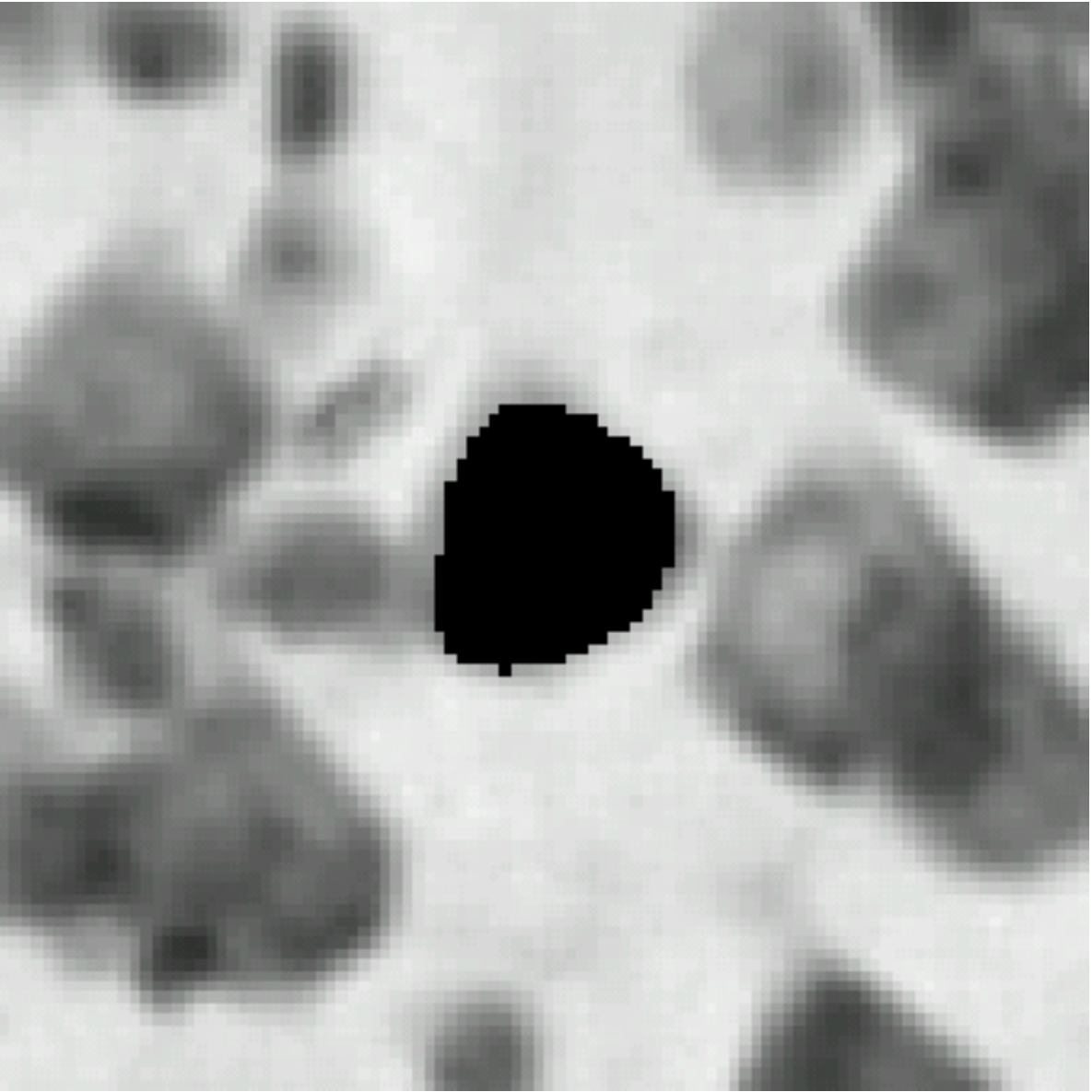} \end{array} \\ 
\mbox{({\bf c})} & \begin{array}{c} \includegraphics*[width=4cm]{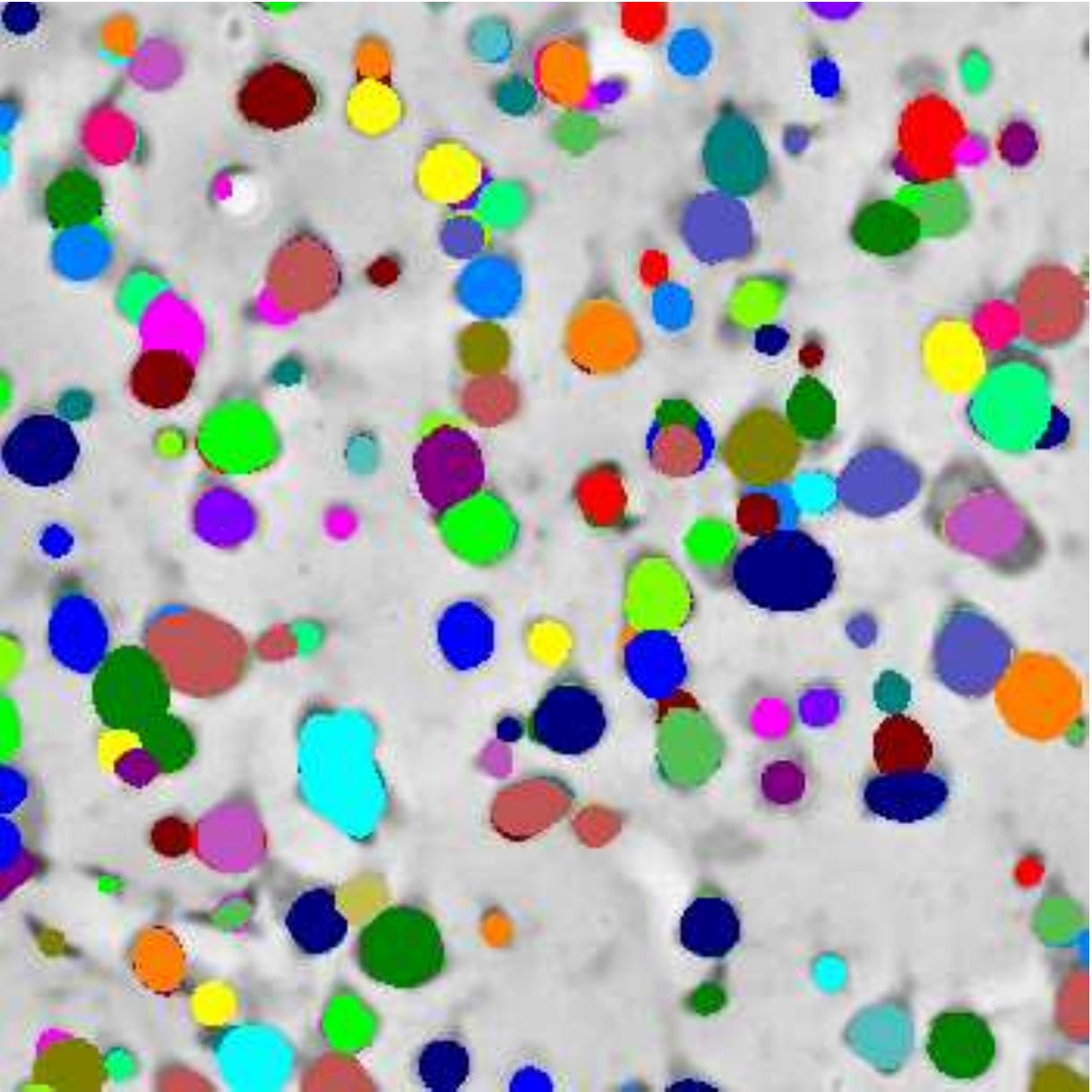} \end{array}
\end{array}$
\end{center}
\caption{Steps of the overall segmentation method (OSM). ({\bf a})
Over-marking the image with a hexagonal grid of points that lay on the
thresholded foreground and center points of a traditional watershed
segmentation. Points within 5 pixels are combined to avoid redundancy. ({\bf
b}) Active contour segmentation: using each starting location found in ({\bf
a}), a segmentation (clustering) process is performed within a small region
of the image to find one possible neuron cell body. This process is then
repeated for each starting location until all starting locations are
exhausted. ({\bf c}) The final set of {\it computer segments}, shown in
different solid colors, is the output of the OSM.}
\label{OSM}
\end{figure}

An initial step of the segmentation process is ``seeding'' the image with one
or more points for each possible neuron cell body. A combination of two
methods is used (Fig.~\ref{OSM}{\bf a}): a hexagonal grid of points is placed over the thresholded
foreground of the image and the center points of objects identified by the
traditional watershed segmentation~\citep{Javi:2002}.

\subsubsection{\label{0b}Active contour segmentation}

We employ active contour segmentation with statistical shape
knowledge~\citep{Cremers:2000} because the method is designed to overcome the
challenges of noisy images and missing boundary data, the main identification
challenge in Nissl-stained tissue. Also, the method uses low-dimensional
shape representations which are ideal for modeling cell contours (outlines of
cells). Because the image is initially over-marked, the calculations of
contour splitting~\citep{Zimmer:2002} are not needed.

The image $f_{ij}$ is a digital image of sliced brain tissue which defines
the optical density (gray scale value) of each pixel ({\it ij}).  We assume
that the image contains at least one type of object of interest (neurons)
mixed with other objects (non-neurons). The goal of a single run of the
segmentation is to ``segment'' a single object of interest (a single neuron)
from the rest of the image (all other neurons, non-neurons, and
background). It does this by ``evolving'' a loop of pixels called a {\it
contour} ($C$) from a circle of typical neuron diameter ($12 {\mu}m$)
starting at one of the over-marked starting points, to a location and shape
that surrounds a potential neuron cell body (Fig.~\ref{OSM}{\bf b}). This
process is repeated for each starting location until all starting locations
have been exhausted.

The movement of $C$ is controlled by a set of $N$ points called {\it control
points} $\left\{\left(x_n,y_n\right)\right\}_{n=1..N}$ for which we use
the compact notation~\citep{Cremers:2000}

\begin{equation}
  {z}=\left({\bf r}_1,...,{\bf r}_N\right)=\left(x_1,y_1,...,x_N,y_N\right)
\end{equation}

\begin{figure}[p]
\begin{center}
$\begin{array}{c c}
\mbox{({\bf a})} & \begin{array}{c} \includegraphics*[width=7cm]{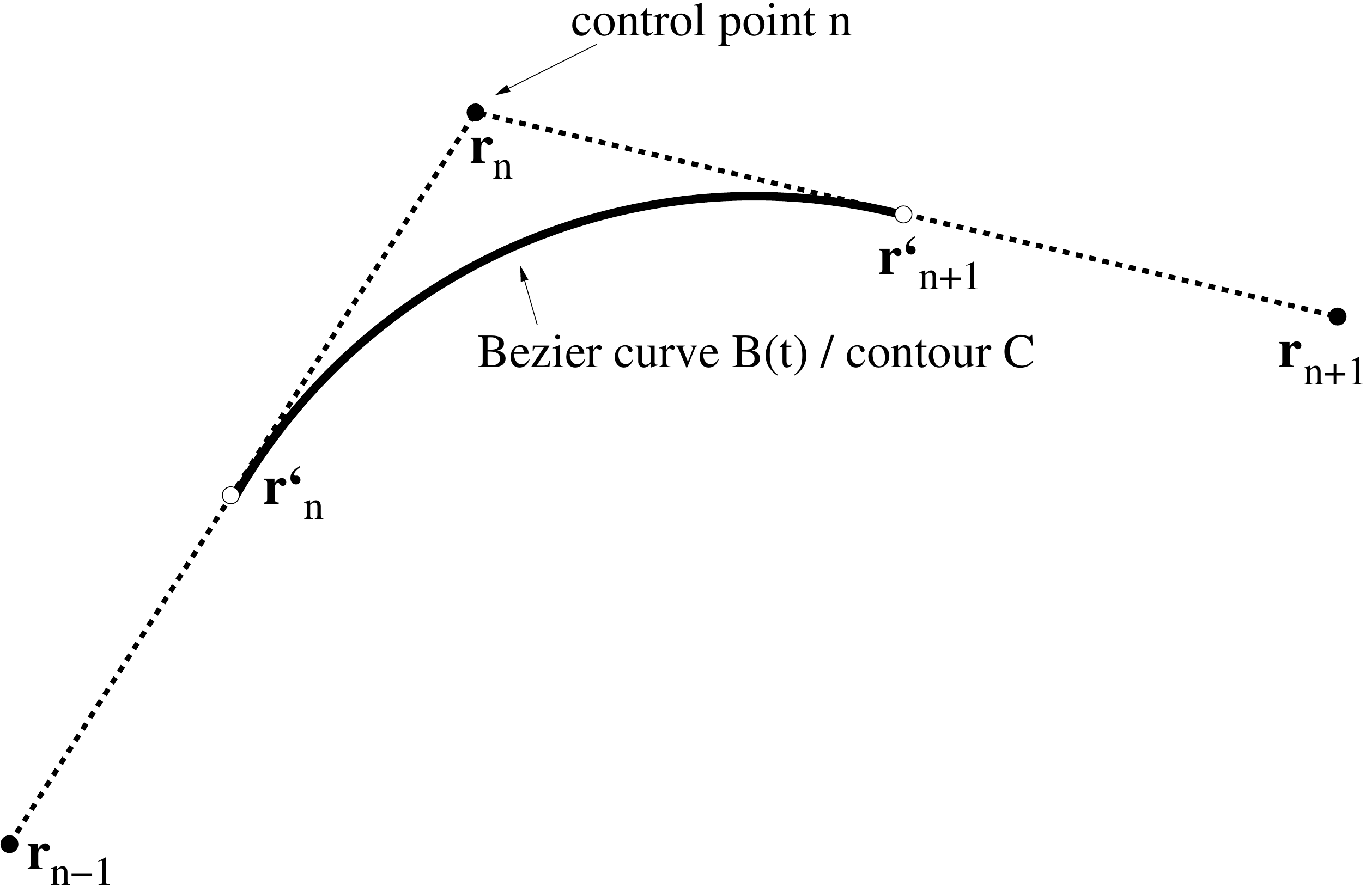} \end{array} \\
\mbox{({\bf b})} & \begin{array}{c} \includegraphics*[width=10cm]{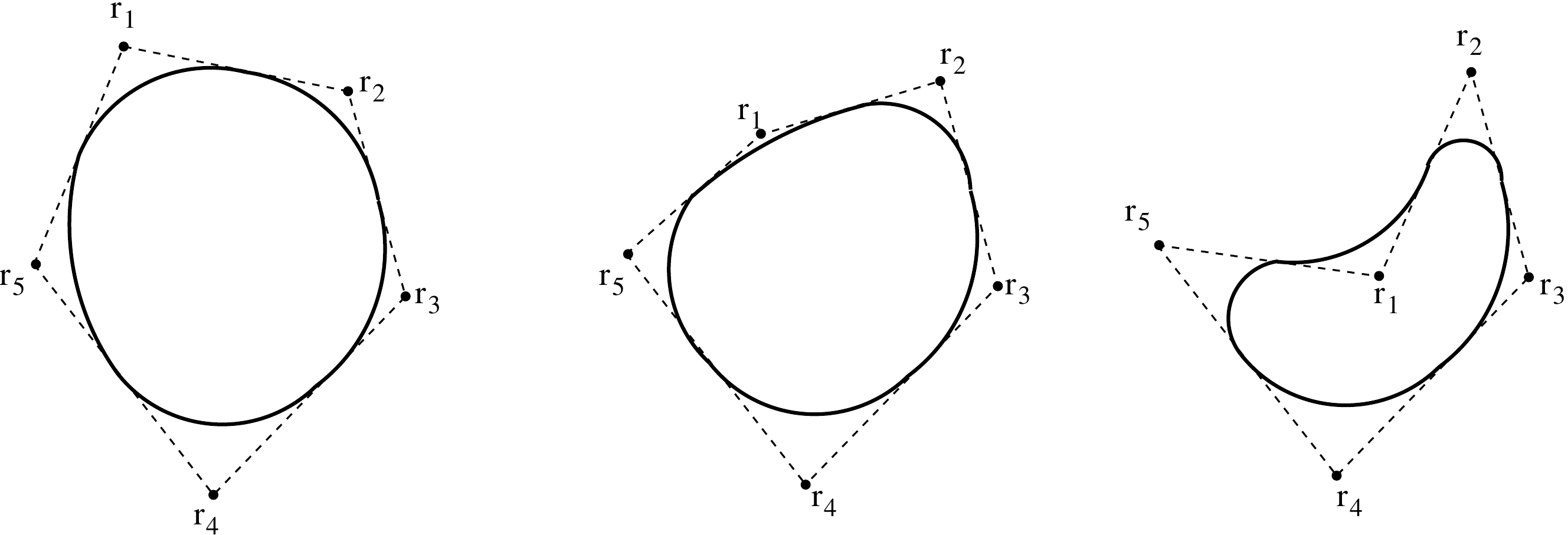}\end{array}  \\
\mbox{({\bf c})} & \begin{array}{c} \includegraphics*[width=5cm]{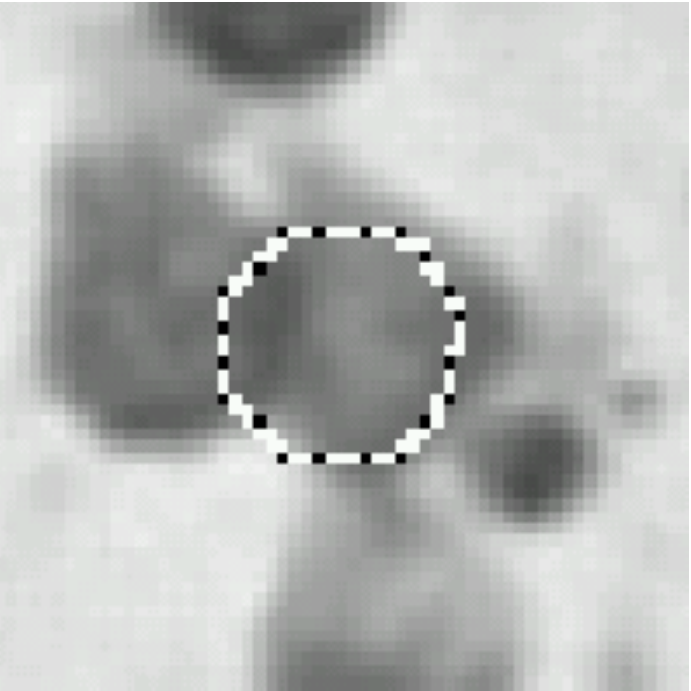}\end{array}  \\
\end{array}$
\end{center}
\caption{ ({\bf a}) The contour $C$ is described by the control points ${\bf
 r}_1,{\bf r}_2 ,...,{\bf r}_N$. ({\bf a}) Quadratic Bezier curve $B(t)$ is
 defined for control point $n$ using the control points ${\bf r}_{n-1}$,
 ${\bf r}_{n}$, and ${\bf r}_{n+1}$. The points ${\bf r}{\prime}_{n}$ are
 halfway between ${\bf r}_{n-1}$ and ${\bf r}_{n}$. The equation for the
 contour is $B(t)=(1-t)^2{\bf r}{\prime}_{n}+2t(1-t){\bf r}_{n}+{\bf
 r}{\prime}_{n+1}t^2,t=0..1$. The equations guarantee that at the points
 ${\bf r}{\prime}$ the curve is continuous and smooth. Combining several
 Quadratic Bezier curves creates a quadratic B-spline contour. An example
 with 5 control points is presented in ({\bf b}) which shows how the B-spline
 contour moves when one control point (${\bf r}_{1}$) moves. ({\bf c})
 Contour C (white pixels) with 20 control points (single black pixels) that
 is overlaying the image.}
\label{control.points}
\end{figure}

\noindent
The control points are parameters in a closed quadratic Bezier-spline
(B-spline) curve~\citep{Blake:1998} that define the exact location (pixels)
of $C$ (see Fig.~\ref{control.points} for definition). Hence, $C$
moves and changes shape by the iterative motion of the control points ${z}$. At
each time step, each control point ${z}$ makes a small movement towards
encircling an object close to its starting location by minimizing a total
energy $E$ based on two energy considerations, $E_{MS}$ and $E_{c}$:

\begin{equation}
E(f,u,C) = E_{MS}(f,u) + {\alpha}E_{c}(C)~\mbox{.}
\end{equation}

\begin{figure}[p]
\begin{center}
$\begin{array}{c c c c}
\mbox{high }E_{MS} & \mbox{low } E_{MS} & \mbox{high } E_{MS} & \mbox{low }E_{MS} \\
\mbox{high }E_{c} & \mbox{high }E_{c} & \mbox{low }E_{c} & \mbox{low }E_{c} \\
(BAD) & (BAD) & (BAD) & (GOOD) \\
\hline
 \includegraphics*[width=3cm]{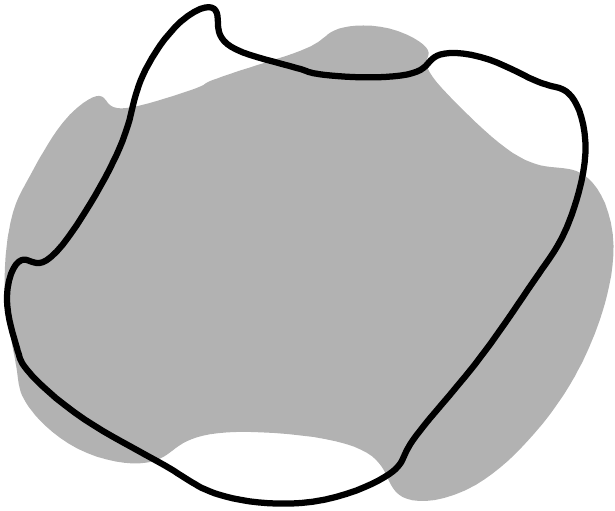} &  \includegraphics*[width=3cm]{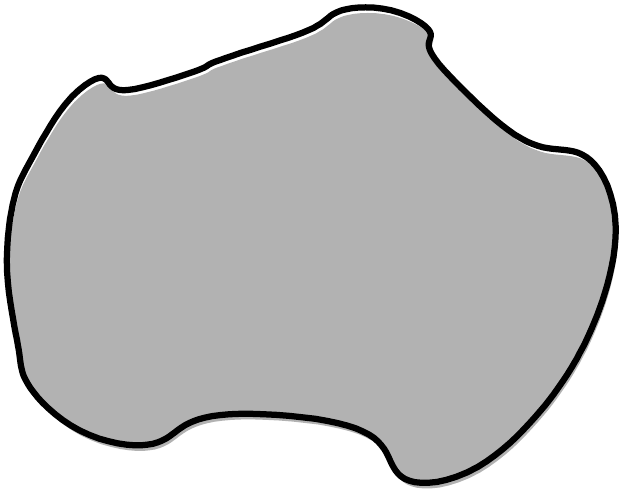} &  \includegraphics*[width=3cm]{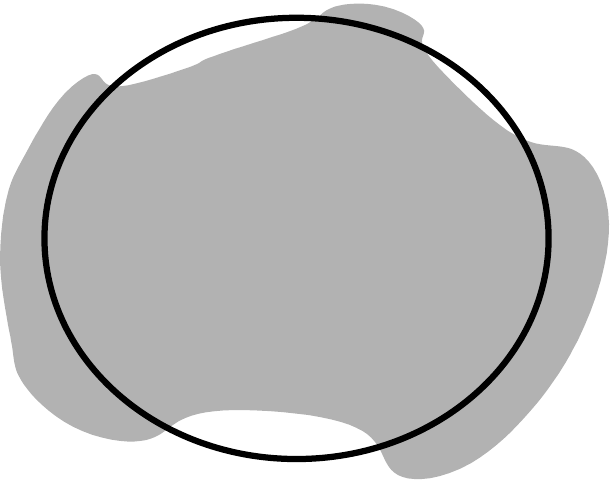} &  \includegraphics*[width=3cm]{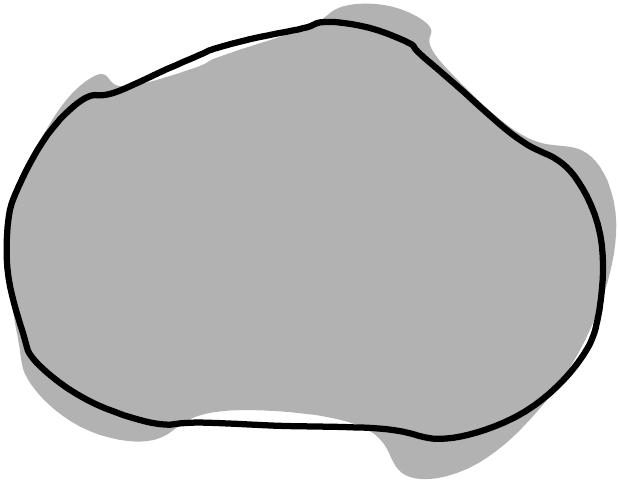} \\
\hline
\end{array}$
\end{center}
\caption{Schematic drawing showing the relative energies of $E_{MS}$ and $E_{c}$ for the same image (shown as gray) and four different contour shapes (shown as black loops). The first three cases are examples of improperly fit contours with a high overall energy $E=E_{MS}+{\alpha}E_c$. The last case is an example of an optimal contour minimizing the overall energy.}
\label{high_low}
\end{figure}

\begin{figure}[p]
  \begin{center}
    $\begin{array}{c c}
      ({\bf a}) & \begin{array}{c} \includegraphics*[width=8cm]{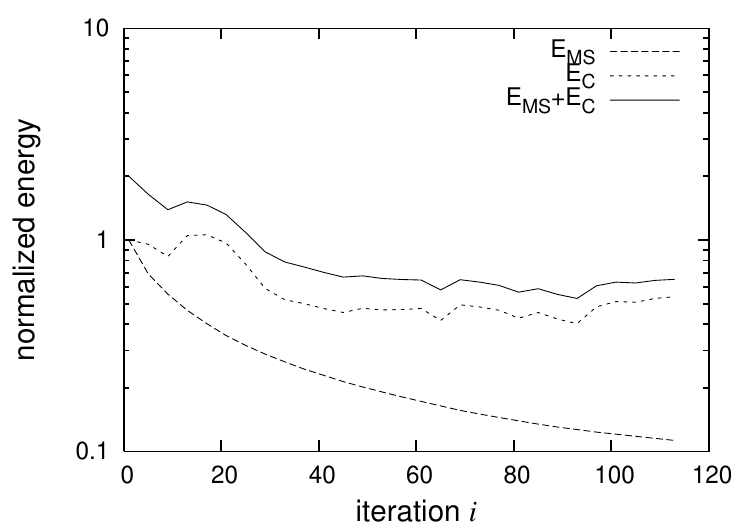} \end{array} \\ 
      ({\bf b}) &\begin{array}{c c c c}
	\includegraphics*[width=2.5cm]{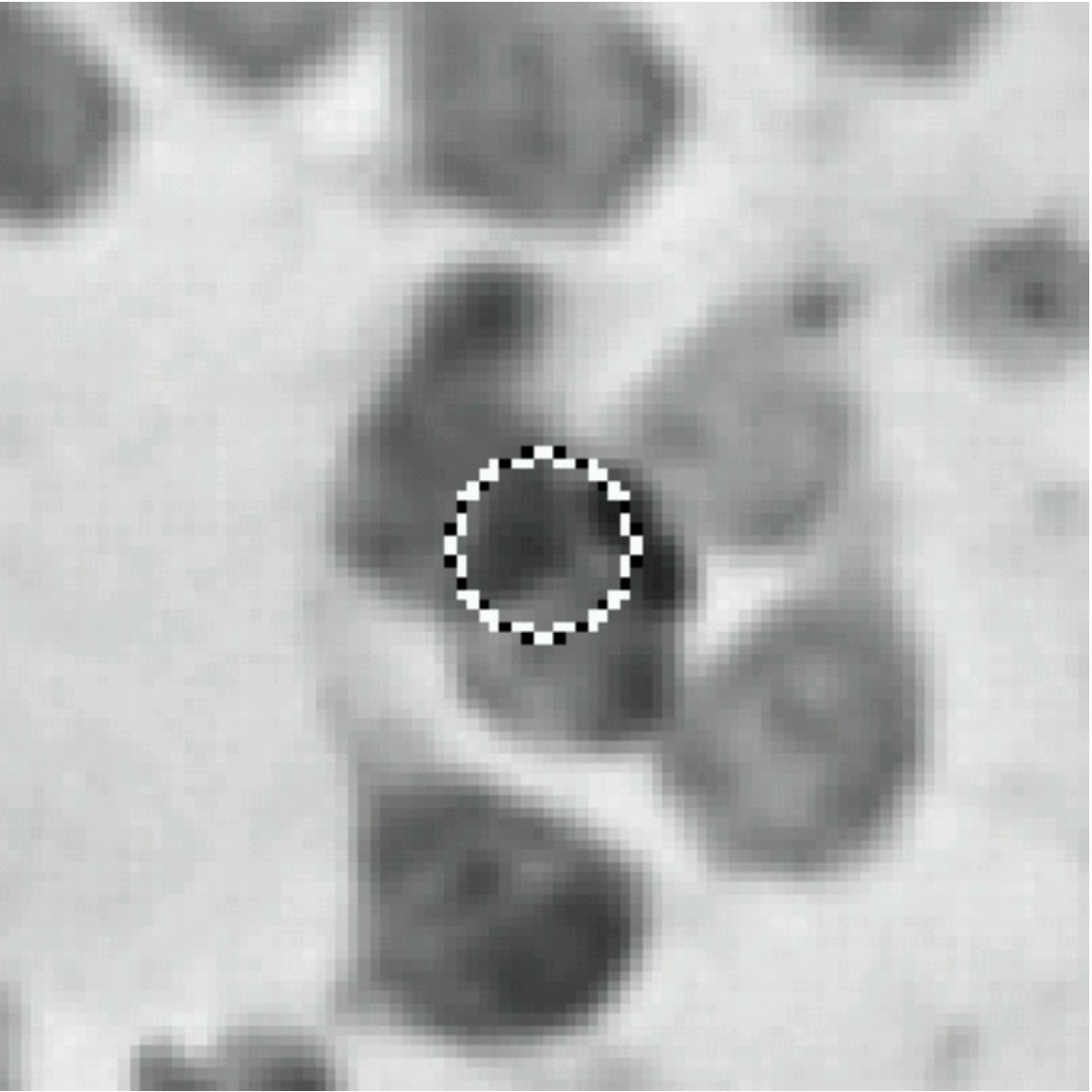} & \includegraphics*[width=2.5cm]{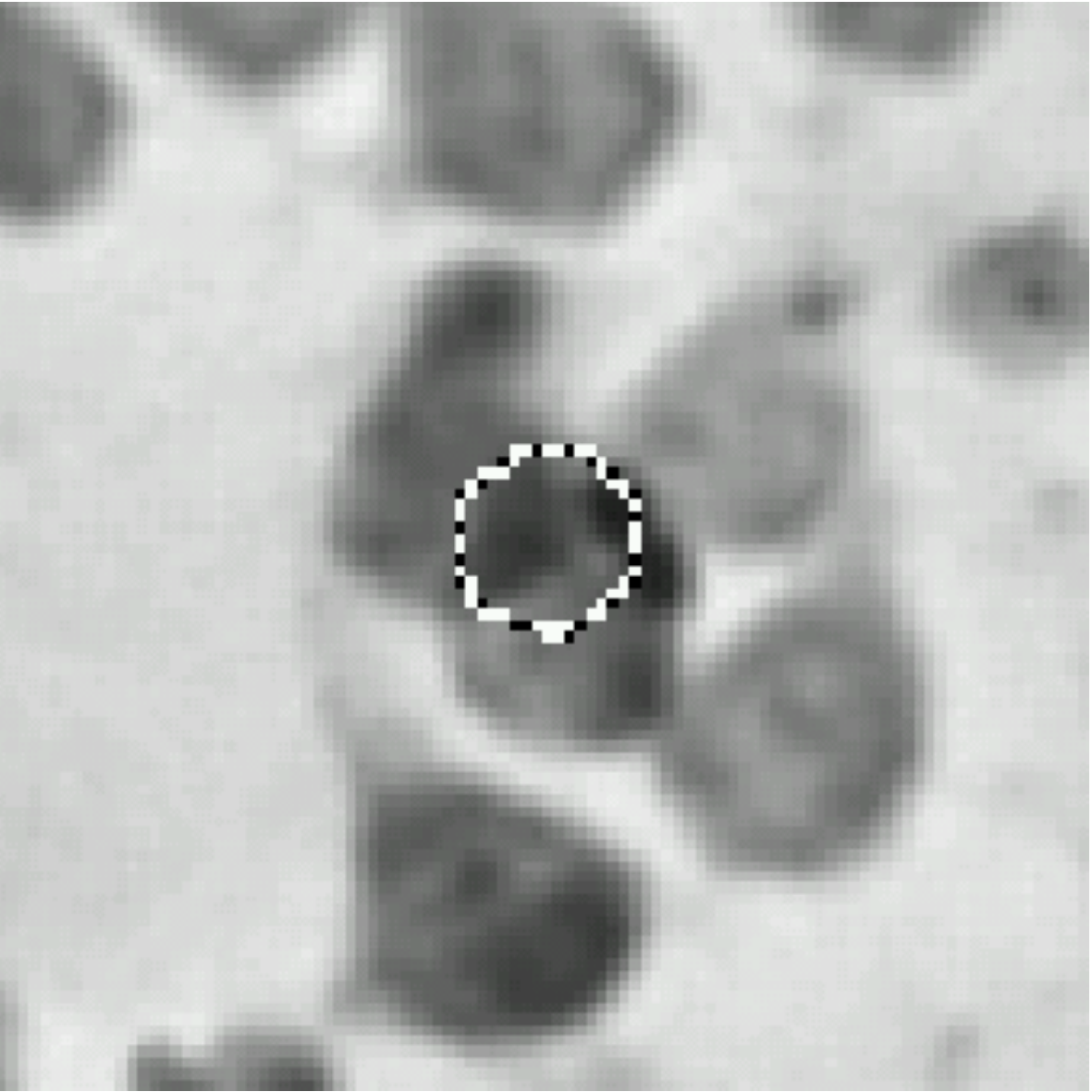} & \includegraphics*[width=2.5cm]{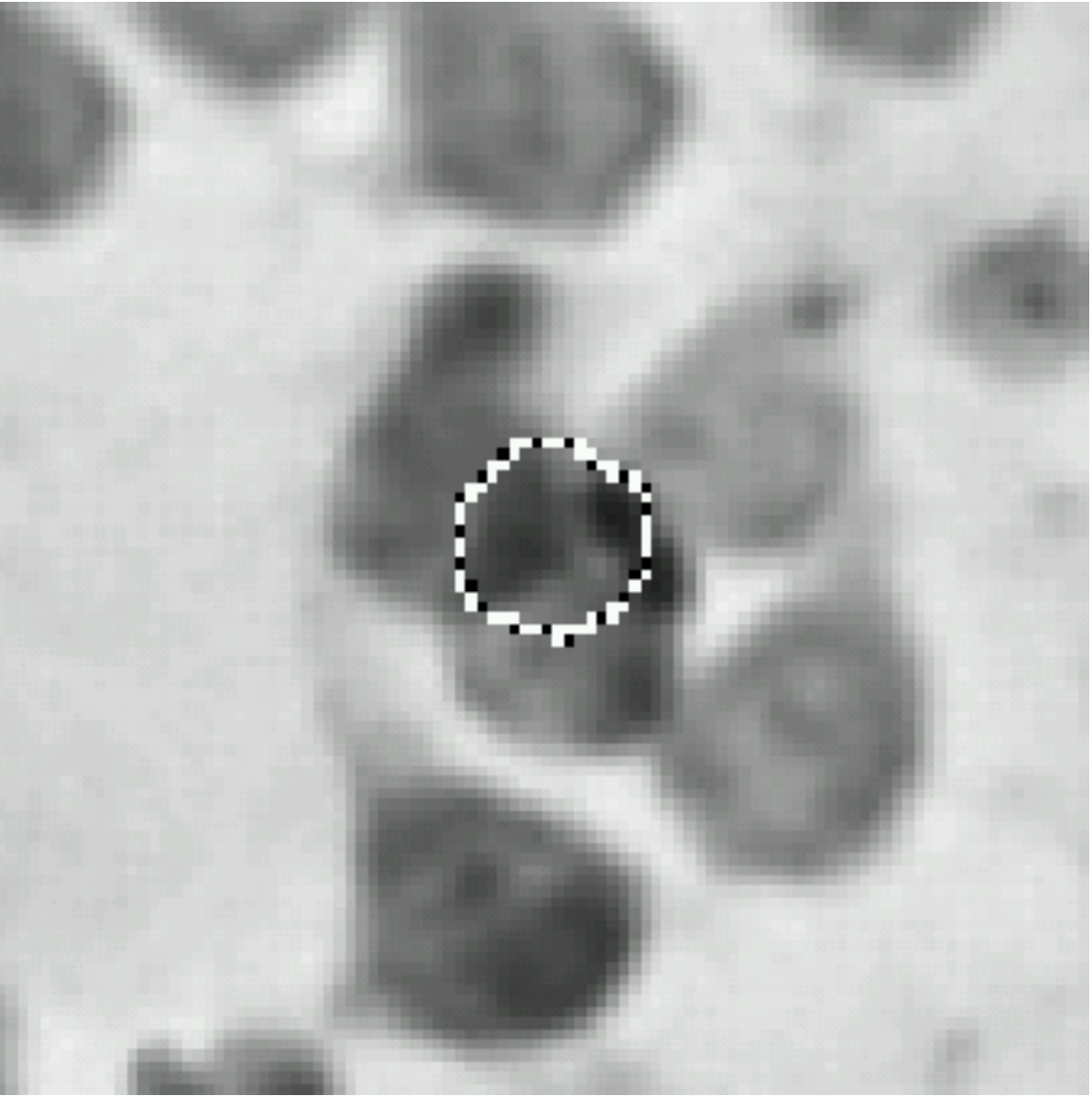} & \includegraphics*[width=2.5cm]{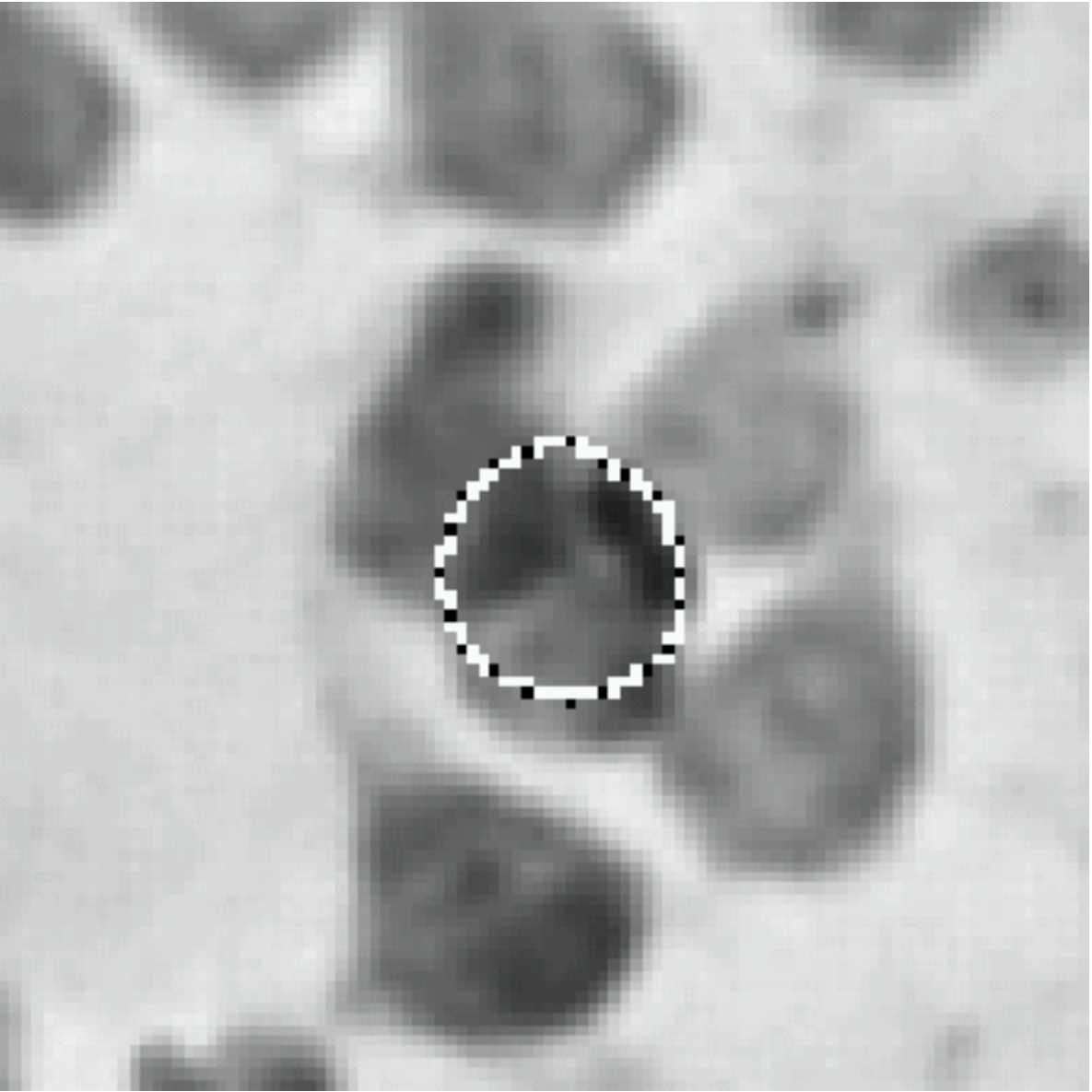} \\
	i=0 & i=4 & i=8 & i=30 \\
	\includegraphics*[width=2.5cm]{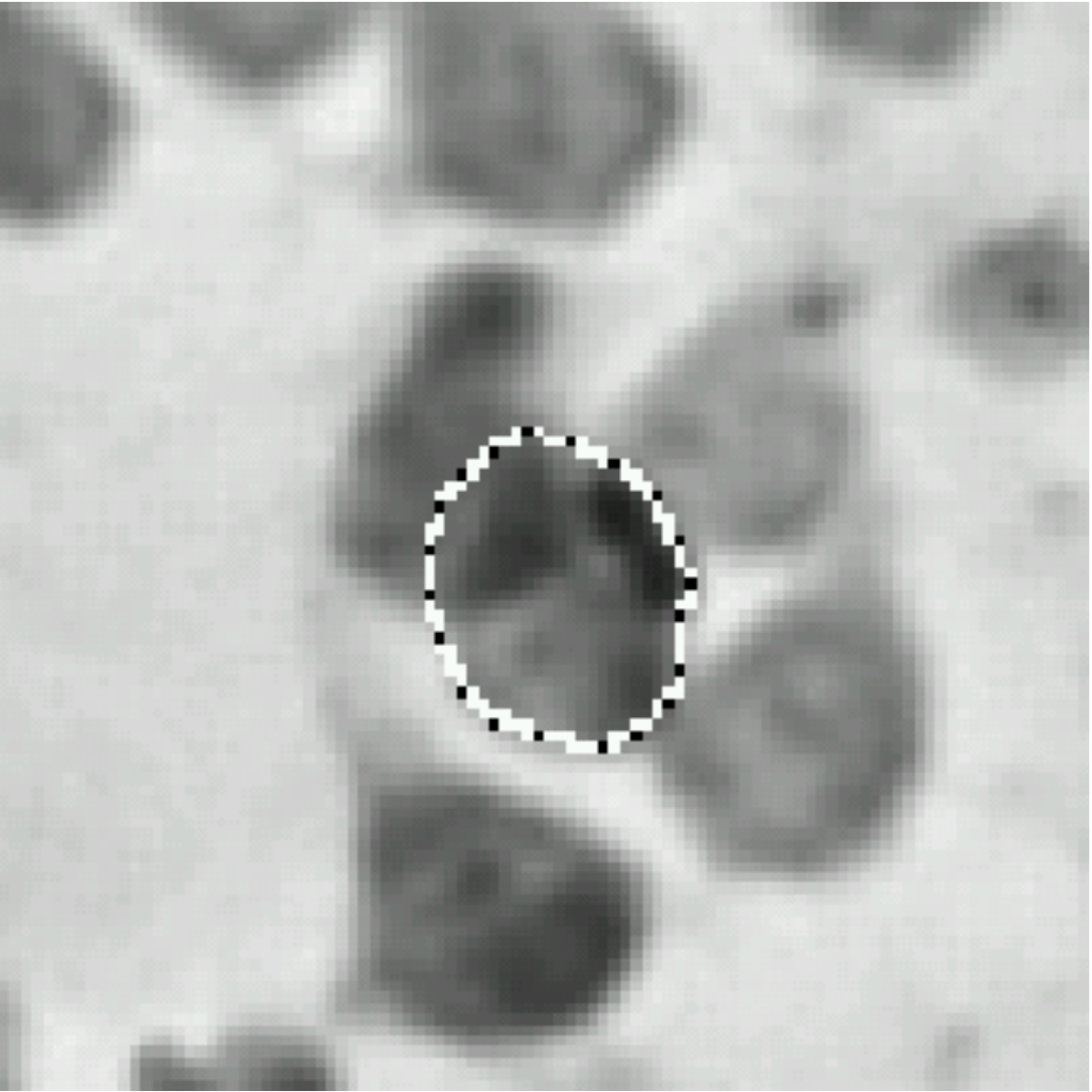} & \includegraphics*[width=2.5cm]{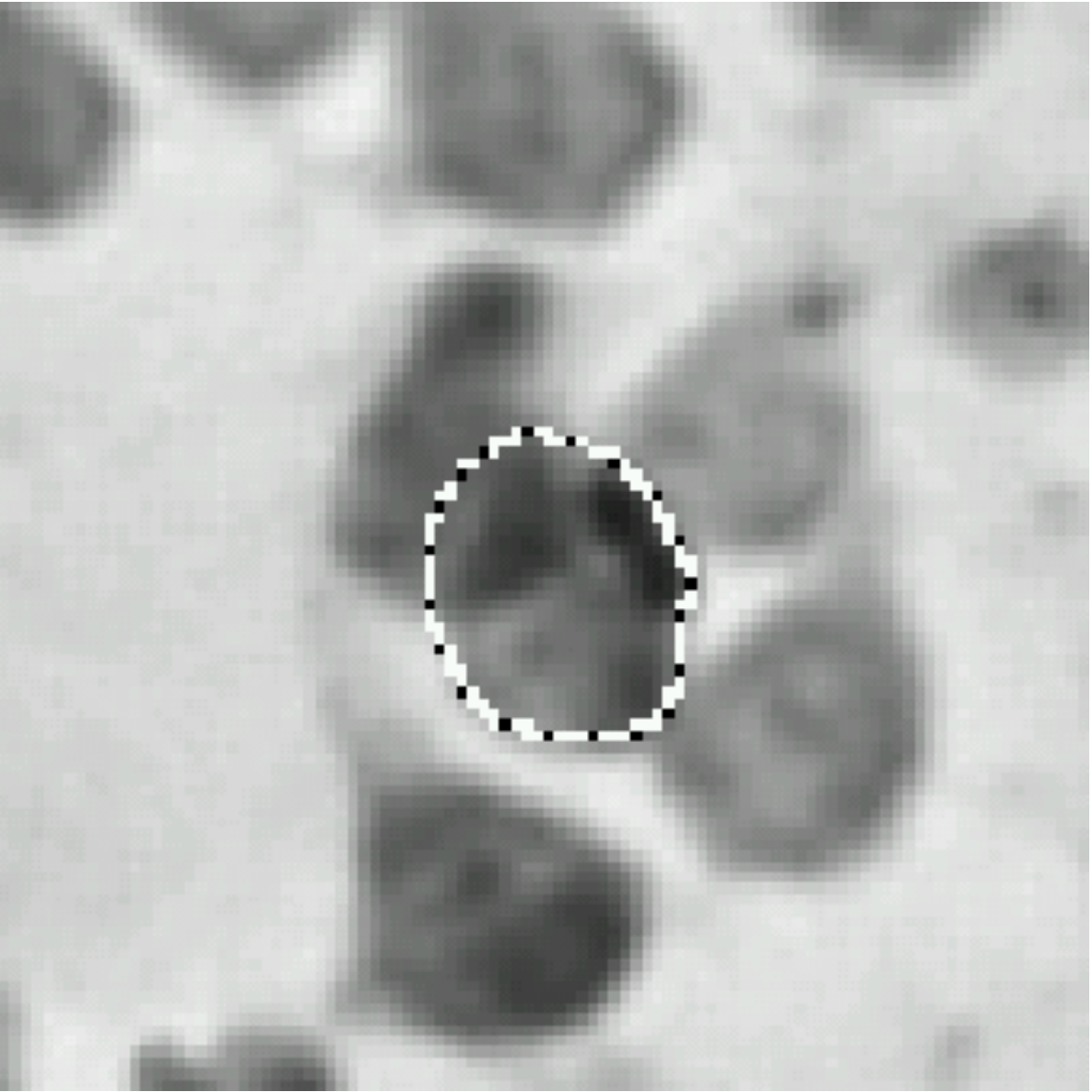} & \includegraphics*[width=2.5cm]{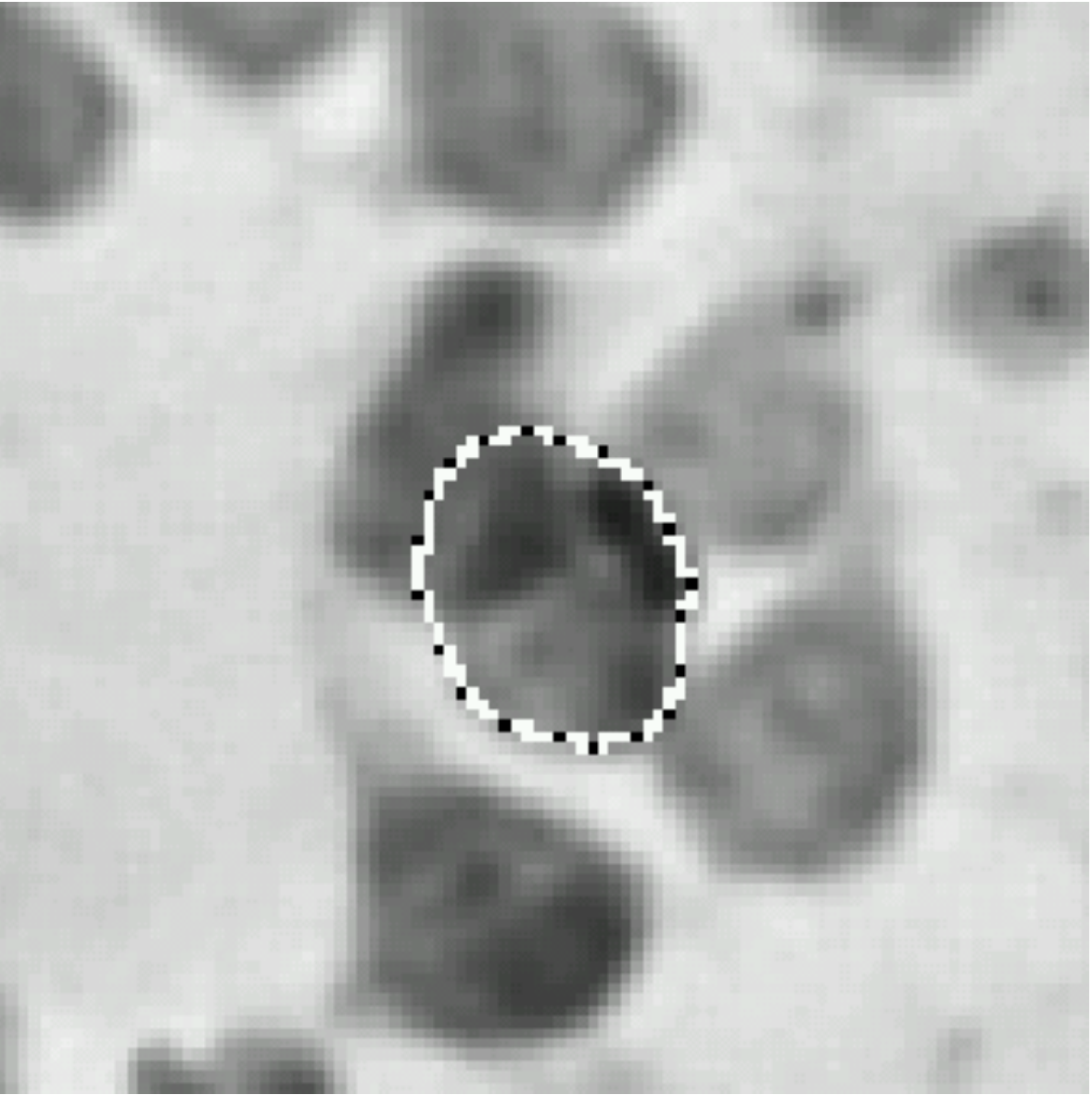} \\
	i=48 & i=60 & i=120 \\
      \end{array}
    \end{array}$
\end{center}
\caption{Examples of the active contour movement within the image during the
OSM segmentation phase. The number of control points (black dots) is 20. The
B-spline contour is white. The contour starts at a location determined by the
over-marking step of the OSM. ({\bf a}) Evolution of the energy terms $E_{MS}$
and $E_{c}$ ({\bf b}) Contour evolution after 0, 4, 8, 30, 48, 60, and 120
steps. When a local minima is reached, the contour no longer moves, and the
points internal to the contour are saved.}
\label{evolution}
\end{figure}

\noindent
A qualitative understanding of the energy terms is presented in
Fig.~\ref{high_low}. $E_{MS}$ is the Mumford-Shah energy term, which
determines how well the contour separates lighter and darker gray scale
regions in the image $f_{ij}$. $E_{c}(C)$ is the contour energy term, which
quantifies the similarity of the contour to a previously chosen set of
training shapes (in our case, the training shapes are oval-like). $E_{MS}$ is
high when $C$ does not separate different contrasts well, and is low if it
does. $E_{c}(C)$ is high if the shape is very contorted, and low if it is
oval-like. $\alpha$ changes the relative influence of the two energy
terms. If $\alpha$ is a high value, then $C$ will evolve into a rigid perfect
oval, ignoring all image information. If $\alpha$ is zero, then $C$ will
surround any nearby object in the image with no regard to the final shape of
$C$. When the two energy terms are balanced with an appropriate $\alpha$ and
the system is evolved to minimize $E$ then objects in an image are encircled
properly.  Fig.~\ref{evolution} shows a typical evolution of $C$ with an
appropriate $\alpha$ value. $u_{ij}$ is a variable image, similar to a
blurred version of $f_{ij}$, which is used in the algorithm, as described
below.

The {\it Mumford-Shah energy term} $E_{MS}(f,u)$ quantifies the alignment of the contour with edges in the image $f_{ij}$:

\begin{equation}
E_{MS}(f,u) = \frac{1}{2}\sum_{ij}\left\{{{ (f_{ij}-u_{ij}(t))^2 + \lambda^2|{\nabla}u_{ij}(t)|^2}}\right\}
\label{ms}
\end{equation}

\noindent
where $\lambda$ is the Mumford-Shah energy parameter that determines relative
strengths of the terms. $|{\nabla}u_{ij}(t)|^2$ is the square of the
magnitude of the picture gradient:

\begin{equation}
{|{\nabla}u_{ij}(t)|^2}={\left(\frac{{\partial}u}{{\partial}x}\right)}^2+
 {\left(\frac{{\partial}u}{{\partial}y}\right)}^2=\frac{\left[u_{i+1,j}(t)-u_{i-1,j}(t)\right]^2+\left[u_{i,j+1}(t)-u_{i,j-1}(t)\right]^2}{4}
\label{grad}
\end{equation}

\noindent
It should be noted that \citet{Cremers:2000} includes an additional term
$\nu\|C\|$ to Eq.~\ref{ms}, which minimizes the length $\|C\|$ of the contour
within its evolution. We do not include this term because it adds an
additional free parameter and does not contribute to the functionality of the
algorithm when identifying cell shaped objects.

\begin{figure}[p]
\begin{center}
$\begin{array}{c}
\includegraphics*[width=7cm]{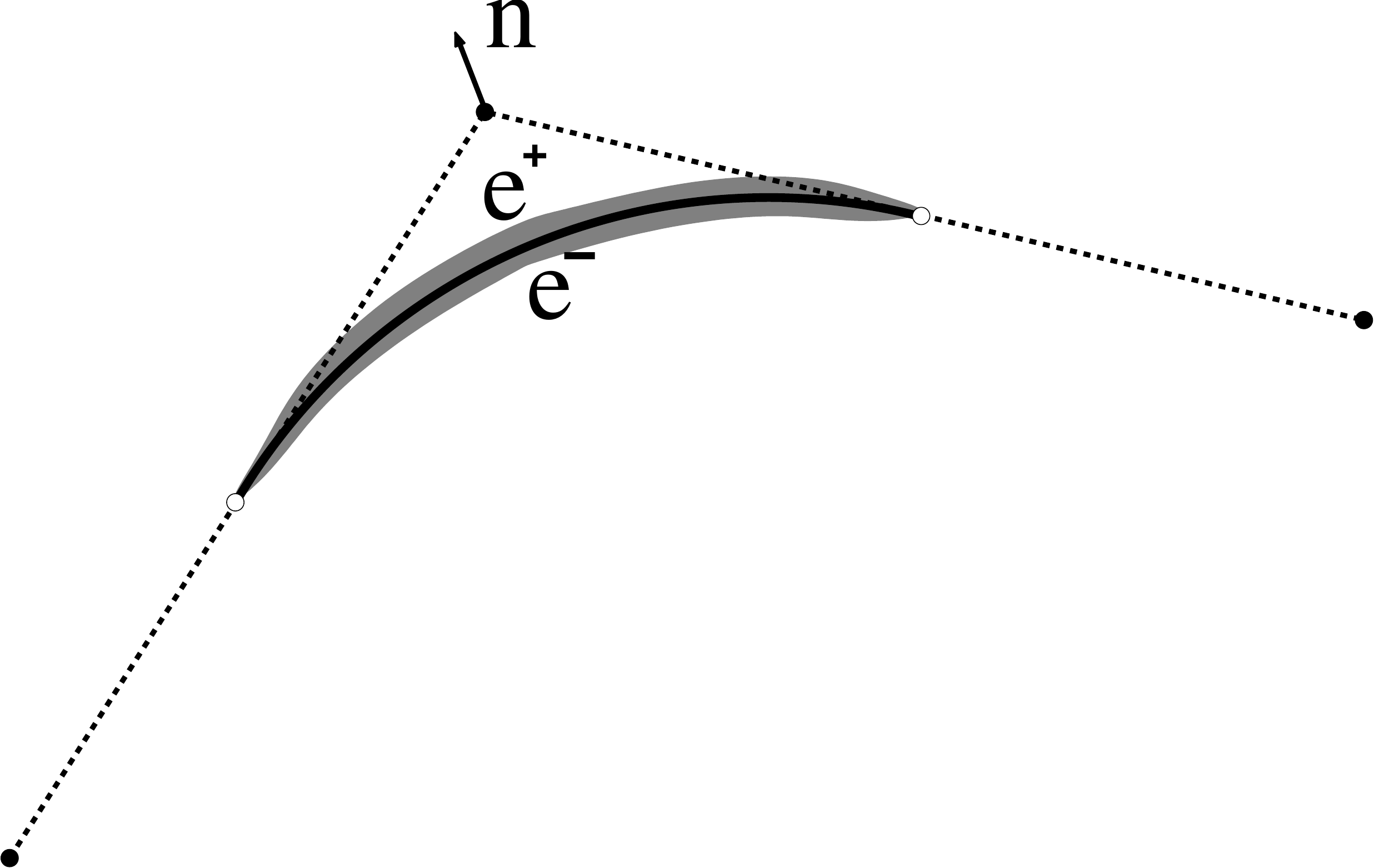} \\
\end{array}$
\end{center}
\caption{Control point movement based on $E_{MS}$ follows Eq.~\ref{both}. The terms $e^{+}$ and
$e^{-}$ are $E_{MS}$ (Eq.~\ref{ms}) integrated over the single line of pixels
right outside ($e^+$) and right inside ($e^-$) of the contour centered around
each control point $n$.  ${\bf n}_x$ and ${\bf n}_y$ are the $x$ and $y$
components of the outer normal vector of $C$ at the control point.}
\label{control.points.2}
\end{figure}

Eq.~\ref{ms} is differentiated with respect to control point
movement. Setting the solution of the differentiation to a minimum of
$E_{MS}(f,u(t))$ gives the evolution equation for each individual control
point $n=1..N$ during each iteration $dt$~\citep{MumfordShah:1989}:

\begin{equation}
 \begin{array}{c}
{\dot{x}}_n\left(t\right) = \left( e^+ - e^- \right){\bf n}_x \\
{\dot{y}}_n\left(t\right) =  \left( e^+ - e^- \right){\bf n}_y\mbox{,}
\end{array}
\label{justms}
\end{equation}

\noindent
where $e^{+}$ and $e^{-}$ are $E_{MS}$ (Eq.~\ref{ms}) summed over the single
line of pixels right outside ($e^+$) and right inside ($e^-$) the segment of
$C$ centered around control point ${\left(x_n,y_n\right)}$ (Fig.~\ref{control.points.2}). ${\bf n}_x$ and ${\bf n}_y$ are the outer
normal vectors of $C$ at each control point ${{\bf r}_n}$ in the x and y
direction respectively. $\dot{x}={dx}/{dt}$ and $\dot{y}={dy}/{dt}$ , where
$t$ is the artificial time parameter. 

Eq.~\ref{ms} is then differentiated with respect to the variable image
$u_{ij}$. Setting the solution to a minimum of $E_{MS}(f,u(t))$ gives the
evolution equation for each pixel $u_{ij}$ during each iteration
$dt$~\citep{MumfordShah:1989}:

\begin{equation}
u_{ij}(t+dt) = 
\left\{ 
  \begin{array}{ll}
    u_{ij}(t)+\left\{{f_{ij}-u_{ij}(t)+\lambda^2{\nabla}^2u_{ij}(t)}\right\}dt & \mbox{if~} ij \ni C \\
    u_{ij}(t) & \mbox{if~} ij \in C 
  \end{array} \right.
\label{justu}
\end{equation} 

\noindent
At $t=0$, $u_{ij}(0)=f_{ij}$. ${\nabla}^2u_{ij}(t)$ is the Laplacian in 2-D Cartesian coordinates:

\begin{equation}
{\nabla}^2u_{ij}(t)=\left(\frac{{\partial}^2u}{{\partial}x^2}\right)+
 \left(\frac{{\partial}^2u}{{\partial}y^2}\right)=u_{i+1,j}+u_{i-1,j}+u_{i,j+1}-u_{i,j-1}-4u_{i,j}
\label{grad2}
\end{equation}

Eq.~\ref{justu} describes a diffusion (${\nabla}^2u_{ij}(t)$) process limited by
the original image ($f_{ij}-u_{ij}(t)$). The key component is that $u_{ij}$
never evolves at the pixels that make up $C$. $u_{ij}$ becomes stable
once $C$ separates contrasted regions. Therefore, minimizing $E_{MS}$ tends to evolve
$C$ so that the gray scale values vary slowly (smoothly) in the
areas inside and outside the contour but vary strongly (discontinuously) across the contour C.

The {\it contour energy term} $E_{c}$ affects the shape of the contour
irrespective of the images $f_{ij}$ and $u_{ij}$. $E_{c}$ is minimized for contour
shapes most similar to a previously chosen set of training shapes
$\chi=\left\{z_1,z_2,...\right\}$. The energy is calculated using the
following equation:

\begin{equation}
E_{c}(C) = \frac{1}{2}\left({{z} - {z_0}}\right)^T{{\it \Sigma}}^{-1}\left({{ z} - {z_0}}\right)~\mbox{,}
\end{equation}

\noindent
where the vector ${z_0}$ and the matrix ${{\it \Sigma}}$ (with an inverse ${{\it \Sigma}}^{-1}$) contain the mean and covariant information of
the previously chosen set of training shapes
$\chi=\left\{z_1,z_2,...\right\}$:

\begin{equation}
  {z_0}=\left<{z_i}\right>
\end{equation}

\begin{equation}
  {{\it \Sigma}}=\left<({z_i}-{z_0})^T({z_i} - {z_o})\right>~\mbox{,}
\end{equation}

\noindent
Here $<>$ denotes the sample average. $z_0$ is a $2N$ vector
and ${\it \Sigma}$ is a $2N$x$2N$ matrix. Creating $z_0$ and ${\it \Sigma}$ for a set of
shapes $\chi=\left\{z_1,z_2,...\right\}$ is equivalent to modeling the
distribution of shapes in ${\mathbb{R}}^{2N}$ as a Gaussian
distribution~\citep{Cremers:2000}.

To minimize $E_{c}(C)$, the following evolution equation for each control point is used:

\begin{equation}
\dot{z}(t)={\it \Sigma}^{-1}\left(z(t)-z_0\right)~\mbox{.}
\label{justc}
\end{equation}

\noindent 

Combining the two equations \ref{justms} and \ref{justc} gives the final
evolution equation for each control point $n$ during each iteration:

\begin{equation}
\begin{array}{c}
{x}_n\left(t+dt\right) =  {x}_n\left(t\right) + \left\{ \left( e^+ - e^- \right){\bf n}_x + \alpha\left[{\it \Sigma}^{-1}\left(z(t)-z_0\right)\right]_{2n-1}\right\}dt \\
{y}_n\left(t+dt\right) =  {y}_n\left(t\right) + \left\{ \left( e^+ - e^- \right){\bf n}_y +  \alpha\left[{\it \Sigma}^{-1}\left(z(t)-z_0\right)\right]_{2n}\right\}dt~\mbox{,}
\end{array}
 \label{both}
\end{equation}

recalling that $e^{+}$ and $e^{-}$ are $E_{MS}$ (Eq.~\ref{ms}) summed over
the single line of pixels right outside ($e^+$) and right inside ($e^-$) the
segment of $C$ centered around control point ${\left(x_n,y_n\right)}$
(Fig.~\ref{control.points.2}), and are dependent on $\lambda$ and $u_{ij}$. 

The evolution of the contour is driven by Eqs.~(\ref{both},\ref{justu}), with
variables $u_{ij}$ and contour points
$\left(x_1,y_1,...,x_N,y_N\right)$. Note that Eqs.~(\ref{both},\ref{justu})
are coupled and must be solved simultaneously.

Performing a step by step evolution of the control points (Eq.~\ref{both})
and $u_{ij}$ (Eq.~\ref{justu}), $C$ evolves in the following way: If $C$
begins to change into a contorted, non-ovular shape to minimize $E_{MS}$
(such as ``leaking'' out of an area of weak or missing boundary information
in the image), then $E_{c}$ will increase, hence there will be a force
opposing the movement. Similarly, if the contour begins to move back to a
perfect oval to minimize $E_{c}$, $E_{MS}$ will increase and thus limit such
a change. When a local minimum is reached and the contour no longer moves,
the points internal to the contour are saved, and the process starts again at
a new location until all starting locations are exhausted.

There are several free parameters ($\alpha$, $\lambda$, $dt$, $N$, etc.) that
must be set within the OSM algorithm. Some of these parameters, called {\it
secondary parameters}, do not greatly affect the evolution, and can be set
the same for all Nissl-stained images. The secondary parameters are as
follows: $N$ is set to 20, so that for a typical 80 ${\mu}m$ circumference of
a neuron cell body, neighboring control points are 3 ${\mu}m$, or roughly 4
pixels away from each other. $z_0$ and ${\it \Sigma}$ define the training
that depend on the typical shapes of the object of interest, in our case a
neuron. We build these parameters by creating a sample of 100 ellipses,
ranging linearly from an eccentricity of 0 to 0.4, a simple representation of
the average shape of neuron cell bodies. To speed up the evolution, we allow
for different ``time'' steps and Mumford-Shah parameters in
Eqs.~(\ref{both},\ref{justu}). In Eq.~\ref{both}, $dt{\rightarrow}dt_{c}$ and
$\lambda{\rightarrow}\lambda_{c}$. In Eq.~\ref{justu},
$dt{\rightarrow}dt_{u}$ and $\lambda{\rightarrow}\lambda_{u}$. In this
schema, $dt_{c}$, $dt_{u}$, and $\lambda_{u}$ can be set as secondary
parameters which do not need to change for any of the pictures. We set
$dt_{c}=100$, $dt_{u}=0.05$, and $\lambda_{u}=1$.

In addition to the secondary parameters, there are two {\it primary
parameters} which greatly affect segmentation, and must be determined
empirically: the energy ratio ${\alpha}$ between $E_{MS}$ and $E_{c}$, and
the energy parameter $\lambda_{c}$ within the $E_{MS}$ term.

Because the active contour algorithm described above was designed for generic
object recognition, the algorithm itself (in addition to the free parameters)
can be ``tuned'' for the task of finding dark elliptical features that are
overlapping or relatively close to each other on a lighter background. We
adjust the above algorithm in a simple way to accommodate overlapping: if
$f_{ij}-u_{ij}(t)>0$ near and inside the given control point, the contour is
``leaking'' out to find the edge of another feature next to it.  We therefore
multiply this control point's contribution to $E_{MS}$ by a free parameter
$\eta$ greater than 1. Here, $\eta$ is a secondary parameter, and is set to
1.5 for all images.

We now discuss each step in ANRA.

\subsection{\label{sec:methods:2}Step I: Image Acquisition}

We test ANRA on Nissl-stained tissue samples of seven young adult (6.4-11.8
years; mean 8.5 years) and seven aged (24.7-32.9 years; mean 30.1 years)
female Rhesus monkey subjects that were part of an ongoing study of the
effects of aging on cognitive function~\citep{Cruz:2004}. For each subject,
eight (4 from each of 2 sections) gray scale (1-256) 512x512 pixel images
with 1.5 pixels/micron resolution ($\sim$150 neurons per image) were taken
from area 46, layer 3 of the prefrontal cortex in the ventral bank of sulcus
principalis. 3 subjects had appreciable differences in image quality between
the two sections, therefore the total number of different
subject/image-qualities is 17. Fig.~\ref{issue_pic}{\bf b} shows 12 of the 17
subject/image-qualities.

\subsection{\label{sec:methods:3}Step II: Segmentation Training}

All images are normalized as described in Sec.~\ref{sec:methods:-1}. Out of
each of the 17 subject/image-qualities, one image is randomly selected as a
{\it training image}. The digital image is marked for neuron cell bodies by
an expert observer who ``paints'' sets of pixels over the neurons using a
small graphical program. Different objects can share pixels, or overlap, but
the sets exist as separate entities even if there is an overlap. We designate
these sets of pixels created by an expert observer the {\it training
segments}. The training segments will be compared to {\it computer segments}
from the OSM output.  The manual identification is relatively quick (2-4
seconds per neuron), and does not require a model image, ie: no feature
overlap~\citep{Lin:2005}. Furthermore, the cell marking method creates
knowledge of the extents of each cell body as viewed by an expert observer,
independent of and unbiased to our segmentation procedure. This information
is saved and used repeatedly for multiple training runs as needed, and
does not have to be repeated for the same image if different training
parameters are checked~\citep{Lin:2005,Lockett:1999}.

We next determine the values of the primary parameters ${\alpha}$ and
${\lambda_{c}}$, the two primary free parameters which greatly affect the
segmentation. We find that there is significant loss in functionality when
${\alpha}$ is outside the $[10^{-9},10^{-8}]$ range and $\lambda_{c}$ is
outside the $[1,5]$ range. We therefore search this space of ${\alpha}$ and
$\lambda_{c}$ by comparing the resulting computer segments to the training
segments. A training segment is ``found'' if the computer segment shares more
than $70\%$ of the pixels with the training segment
(Fig.~\ref{overlap_example}). The set called the final OSM parameters,
denoted ${\alpha}^*$ and ${\lambda_{c}}^*$, is the set that correctly
identifies $95\%$ or more training segments. The
(${\alpha}^*,{\lambda_{c}}^*$) values are then recorded and used for the rest
of ANRA.

The OSM with the correct primary parameters (${\alpha}^*,{\lambda_{c}}^*$)
identifies 95\% or more of neurons in the images, but it also identifies
other non-neuron objects, such as staining errors, glial cells, and improper
coverings of neurons. To separate neurons from non-neurons, computer training
is performed.

First, we compare the (${\alpha}^*,{\lambda_{c}}^*$)-parameter OSM computer
segments to the training segments. Each computer segment is either placed in
the neuron segment category or non-neuron segment category based on whether
the segment mutually overlaps any training segment (Fig.~\ref{overlap_example}). Second, each segment is
represented by seven {\it segment properties} ${\bf
v}=(v_1,v_2,...,v_7)$. The seven segment properties were chosen to be the
most salient measures of identifying neurons within an image. For the
calculations of the segment properties, we denote the total number of pixels
within the segment as $A_c$ and the total number of pixels within the contour
as $\left|C\right|$. The properties are based on the optical density of the
original image $f_{ij}$ as well as the square of the magnitude of the image
gradient $\left|{\nabla}f_{ij}\right|^2$. The segment properties are
presented in Table I. $\sum^{A}$ is a sum over all of the pixels within the
segment area, $\sum^{C}$ is a sum over the edge pixels of the segment
circumference, $r_{c}$ is the location of the center of the segment, and
$r_{ij}$ is the location of the pixel ($ij$).

\begin{figure}[p]
\begin{center}
$\begin{array}[c]{c}
\includegraphics*[width=7cm]{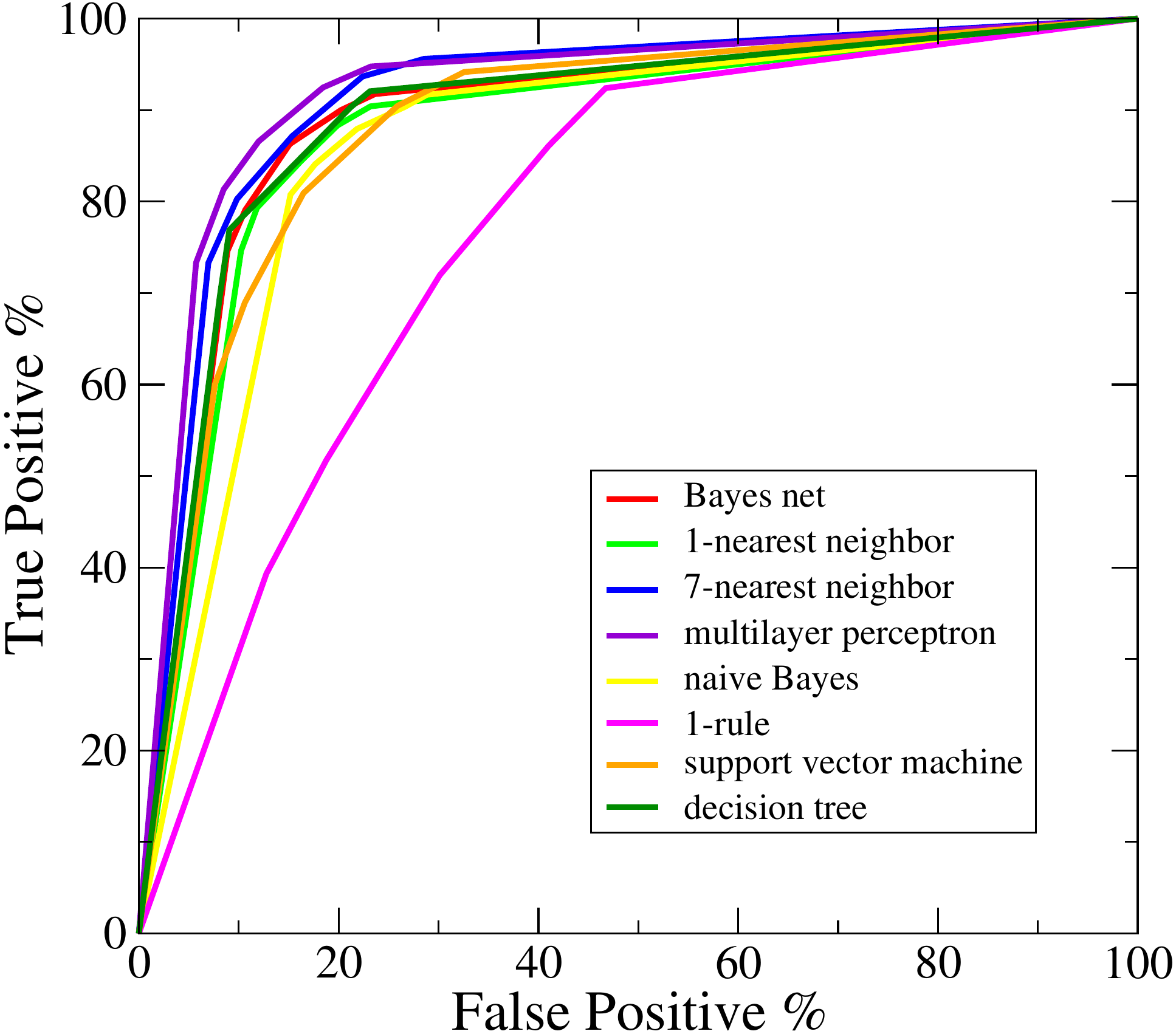}
\end{array}$
\end{center}
\caption{Receiver operating characteristic (ROC) curve for each training
method evaluated. It is seen that the Multilayer Perceptron (MLP) has the
best ROC curve - the highest percentage of neuron property vectors
identified with the smallest percentage of non-neuron property vectors
incorrectly identified. MLP is chosen as the main training method for ANRA. }
\label{ROC1}
\end{figure}

Using the WEKA machine learning toolkit~\citep{Witten:2005}, we assess the
following machine learning algorithm's ability to discriminate between neuron
property vectors $\left\{{\bf v}^+_1,{\bf v}^+_2,...\right\}$ and non-neuron
property vectors $\left\{{\bf v}^-_1,{\bf v}^-_2,...\right\}$: the 1-rule
classifier~\citep{Holte:1993}, naive Bayes classifier~\citep{John:1995},
support vector machine classifier~\citep{Platt:1998}, nearest neighbor
classifier~\citep{Aha:1991}, decision tree classifiers~\citep{Quinlan:1993},
Bayes net and multi-layer perceptron~\citep{Witten:2005}. The cost between
Type 1 errors (marking a non-neuron property vector as a neuron) and Type 2
errors (marking a neuron property vector as a non-neuron) is scanned by
tuning the cost ratio term in the training algorithm. A stratified
cross-validation evaluation for various cost ratios (3:1,2:1,...,1:3) creates
a receiver operator characteristic (ROC) curve~\citep{Duda:2001} for each
training method (Fig.~\ref{ROC1}). The Multilayer Perceptron (MLP) using a
single, 4-node hidden layer, has the best ROC curve, as it provides the
highest percentage of neuron property vectors identified and the smallest
percentage of non-neuron property vectors incorrectly identified. MLP is
therefore chosen as the main training method for ANRA.

\subsection{\label{sec:app}Step III: Application}

Automatic neuron recognition is now applied on an unlimited number of other
images that are normalized and similar in morphology to the training
images. The steps are as follows:

\begin{enumerate}

\item

The OSM with the primary parameters  (${\alpha}^*,{\lambda_{c}}^*$) is performed on the new image.

\item

The properties {\bf v} are calculated for each computer segment.

\item

A cost ratio is selected by the user.

\item

All computer segments deemed non-neurons by the MLP are discarded.

\item

For any two remaining computer segments that mutually overlap by more than
70\%, the computer segments with the smaller probability of being a neuron (as
determined by the MLP) is discarded.

\end{enumerate}

The (x,y) centers, sizes, and shapes of the remaining computer segments are the final result of ANRA.

\subsection{\label{sec:comparisons}Comparison method}

A semi-automatic method (semi-auto) was used in prior neuron density maps
correlation studies~\citep{Cruz:2005}. In the semi-auto method a combination
of computer software and human intervention for each image is employed to
identify neurons. Because the amount of human intervention scales with the
number of images analyzed, the semi-auto method represents a standard with
which we evaluate our completely automated recognition method.

\subsection{\label{sec:dmm}Density Map Method and Microcolumnar Strength}

We give a description of the density map method, as it is the main analysis
to be applied to the results of ANRA. The density map method was initially
described by \citet{Buldyrev:2000} and a more detailed description and
validation was given by \citet{Cruz:2005}. The density map is a 2D
representation of the density correlation function g(x,y), which uses as input
the (x,y) locations of all neurons in the region of interest (ROI). This
function $g(x,y)$ is mapped to a two-dimensional gray scale image (density map) in which
different shades of gray are proportional to the average local neuronal
density. Thus, the density map quantifies the average neuronal neighborhood
surrounding a typical neuron within the ROI.

Operationally, the density map is calculated by first assigning indices
($i=1,2,3...N$) to all the neurons in the sample. Next, we center a grid of
bins of size $D$ over each neuron and count how many other neurons fall in
each bin constructing one matrix of accumulated neurons $m(x,y)$. We define
$g(x,y) = m(x,y) / N{\cdot}D{\cdot}2$, in which $g(x,y)$ has units of an
average density of objects at position $(x,y)$. As an example, the density
map would be uniform if locations of objects (neurons) are uncorrelated, but
will show patterns when there are regular spatial arrangements between
the objects.

For the case of neurons forming microcolumns, their density map exhibits one
central vertical ridge, sometimes accompanied by two less pronounced parallel
neighboring ridges. For this study, we are interested in the microcolumn
strength $S$, which is extracted from the density map by taking the ratio of
the neuronal density within the average microcolumn to the average neuronal
density~\citep{Cruz:2005}. For the same images, $S$ is calculated using ANRA
$(x,y)$ locations as well as semi-automatic $(x,y)$ locations, and the
results are compared.

\section{\label{sec:evaluaiotn}Results}

\begin{figure}[p]
\begin{center}
$\begin{array}{c}
\includegraphics*[width=7cm]{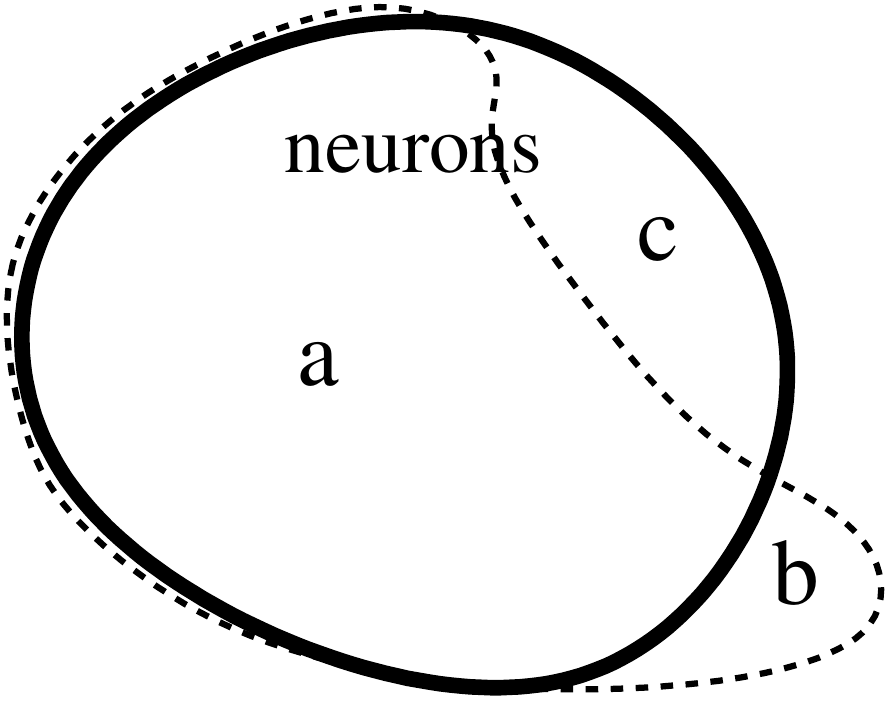} \\
\end{array}$
\end{center}
\caption{Venn diagram showing the relative quantities for evaluating the
quality of a neural recognition method. The bold black line separates neuron
from non-neuron objects in the image. The dotted area shows the objects that
are identified by a method. The method correctly identifies most of the
neurons ({\bf a}), but misses some neurons ({\bf c}) and identifies some
non-neurons as neruons ({\bf b}). Using the quantities {\bf a},{\bf b}, and
{\bf c}, standardized percentages of neuron vs. non-neurons can be
calculated.}
\label{percentages}
\end{figure}

For each of 17 subject/image-qualities, an {\it evaluation image} is randomly
selected from the remaining images and marked for neuron cell bodies by the
expert. The evaluation image is used as a ``gold standard'' to assess the
accuracy of ANRA and the comparison methods. A total of 2448 ``gold
standard'' neurons are analyzed, for an average of 144 neurons per
subject/image-quality. For each of the two recognition methods (semi-auto and
ANRA), we compare the method's identified neurons to the ``gold standard'', and retrieve the following numbers (Fig.~\ref{percentages}):

\begin{equation}
a=\mbox{number of correctly identified neurons}~\mbox{,}
\end{equation}
\begin{equation}
b=\mbox{number of non-neurons incorrectly identified as neurons}~\mbox{,}
\end{equation}
and
\begin{equation}
c=\mbox{number of non-identified neurons}~\mbox{.}
\end{equation}

\noindent
To compare methods for the different subject/image-qualities, we define the
following normalized metrics:

\begin{equation}
A=\frac{a}{a+c}\cdot100~\mbox{,}
\end{equation}
and
\begin{equation}
B=\frac{b}{a+c}\cdot100~\mbox{.}
\end{equation}

\smallskip 

$A$ is the percent of correctly identified neurons (``true positives''). $B$ is the percentage of non-neurons that are incorrectly identified as neurons (``false positives'').

\begin{figure}[p]
\begin{center}
$\begin{array}{c}
\includegraphics*[width=7cm]{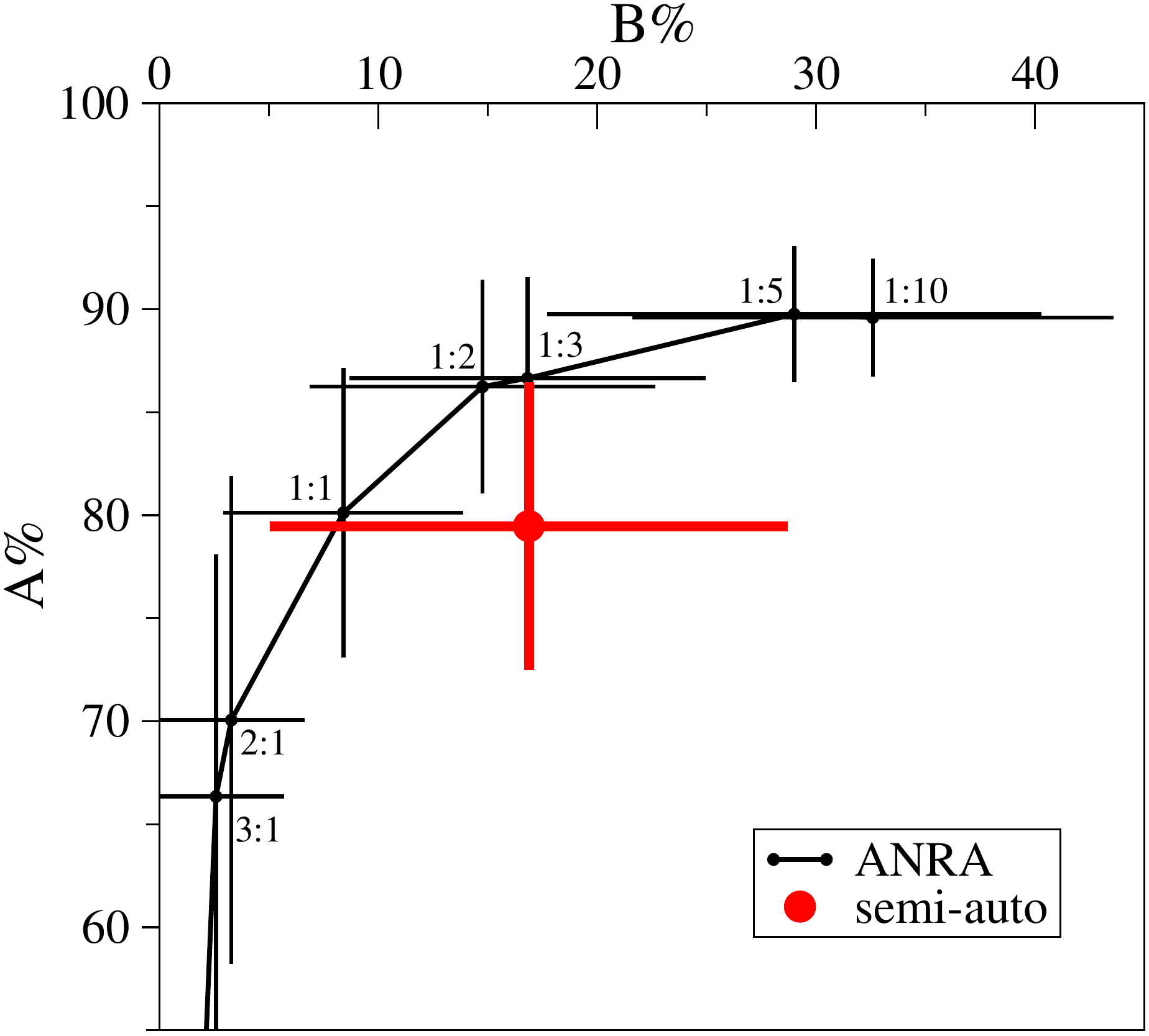} \\
\end{array}$
\end{center}
\caption{Results of ANRA. The semi-auto method is characterized by one
($A,B$) set. Becasue of the ability to adapt the cost ratio
as described in Sec.~\ref{sec:app}, ANRA is shown at 7 different ratios (3:1,
2:1, 1:1, 1:2, 1:3, 1:5, and 1:10), creating an ``adapted'' ROC curve. Since
each point is an average of the 17 subject/image-types, the error bars show
the standard deviation of the spread for both A and B.  }
\label{ROC2}
\end{figure}

\begin{figure}[p]
\begin{center}
$\begin{array}{c c}
\mbox{({\bf a})} & \begin{array}{c} \includegraphics*[width=7cm]{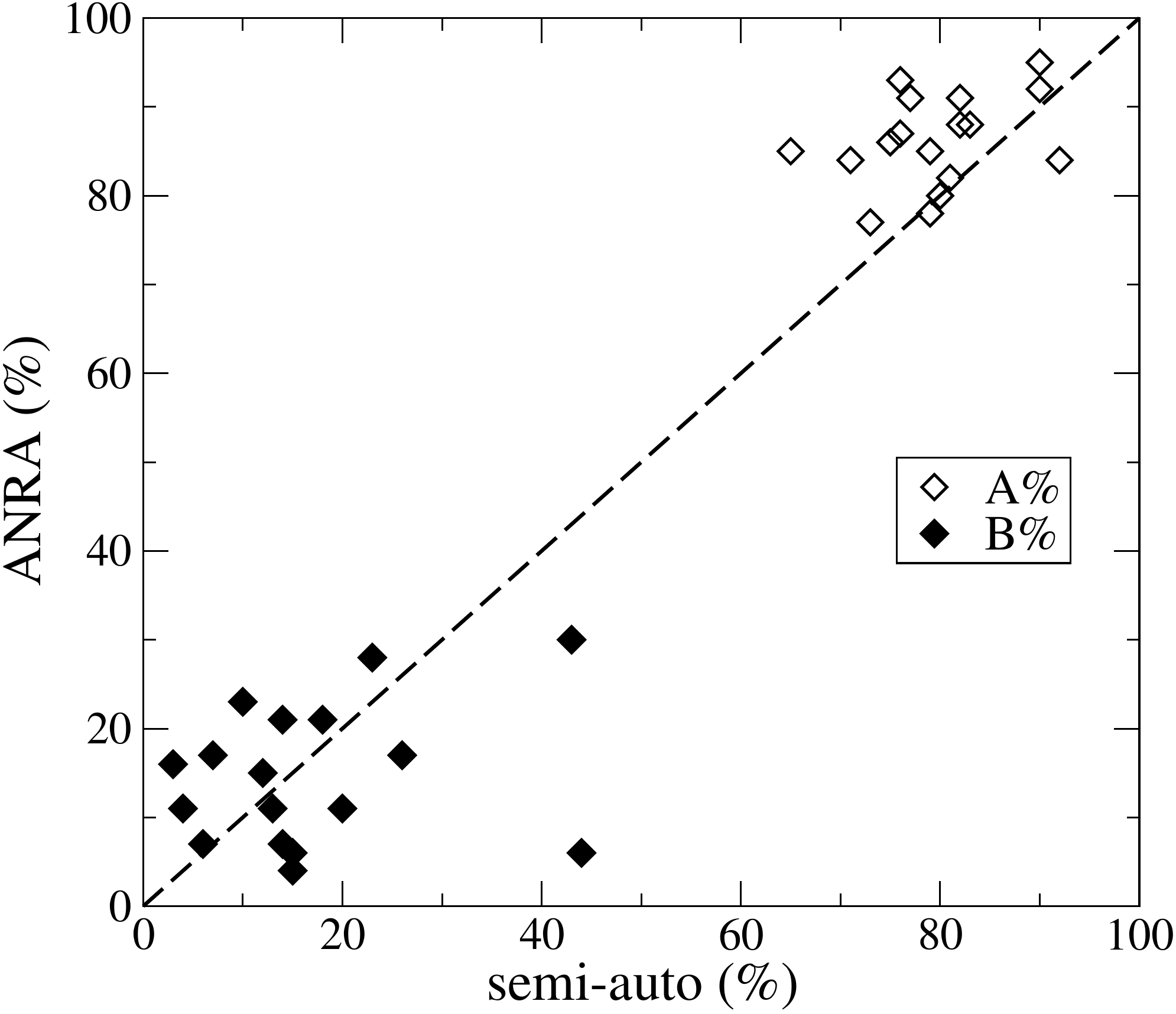} \end{array}\\
\mbox{({\bf b})} & \begin{array}{c c} \includegraphics*[width=6cm]{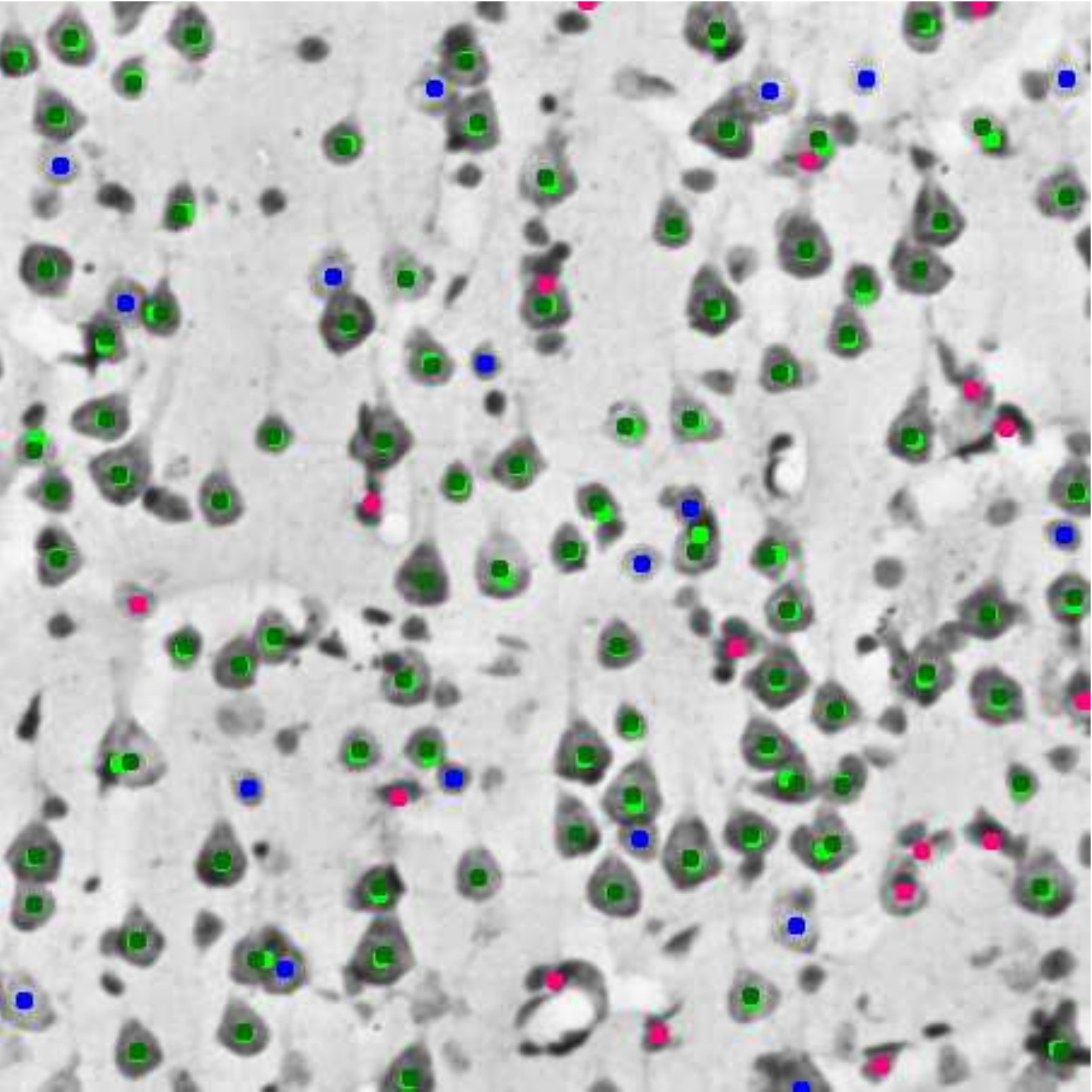}~\includegraphics*[width=6cm]{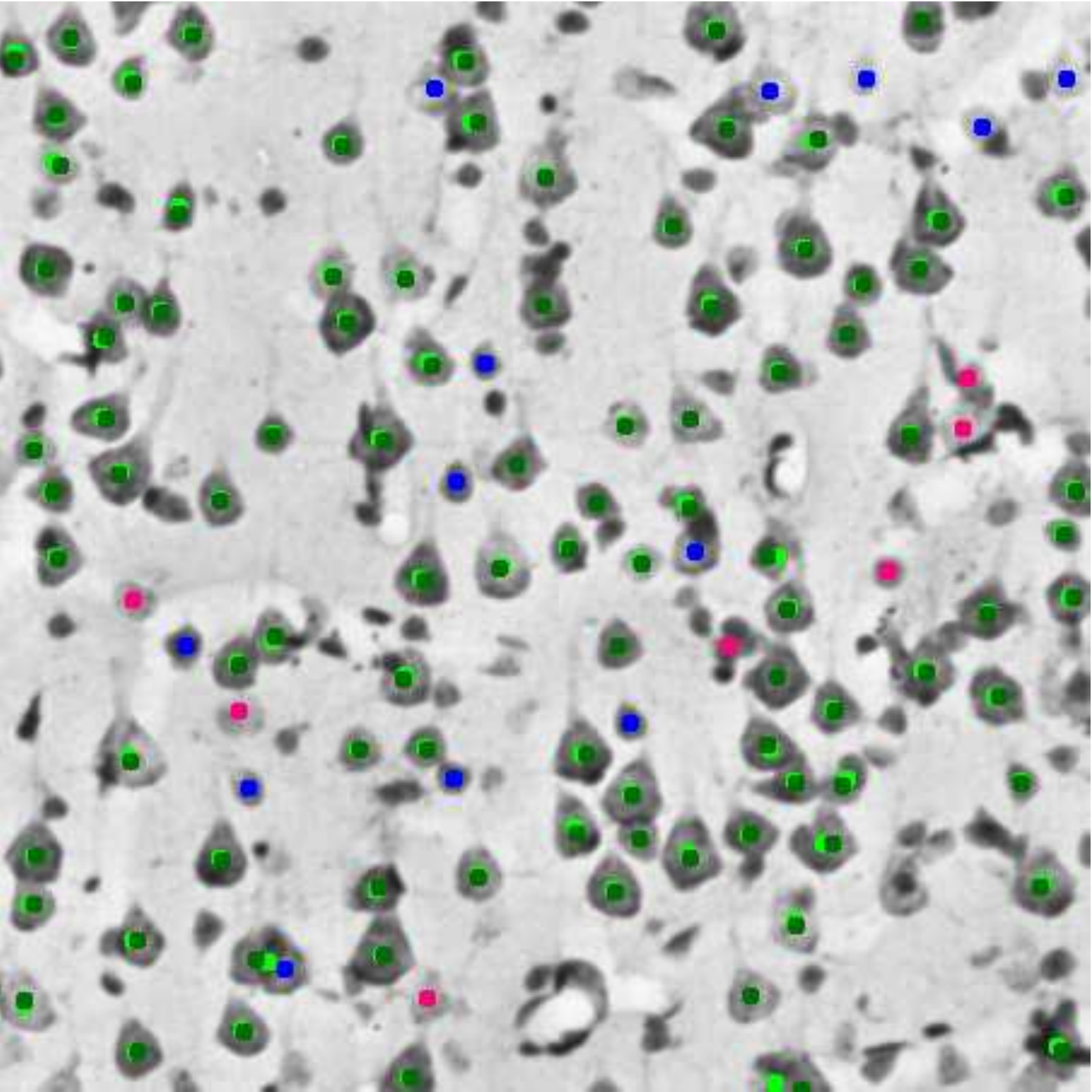}\\
\end{array}
\end{array}$
\end{center}
\caption{({\bf a}) Individual results for 17 subject/image-types for the
semi-auto method and the ANRA (with 1:2 cost ratio). ({\bf b}) Recognition
results for the semi-auto method (left) and the ANRA method (right) for
example subject/image-quality \#1 (Table II). Dark green: gold standard marks
that match with the method.Blue: gold standard marks that DO NOT match with
the method. Light Green: method points that match with gold standard
points. Pink: method points that do not match with gold standard
points.}
\label{Table2graph}
\end{figure}

The results are shown in Fig.~\ref{ROC2}. The semi-auto method is
characterized by one ($A,B$) set. Because of the ability to adapt the cost
ratio as described in Sec.~\ref{sec:app}, ANRA is shown at 7 different ratios
(3:1, 2:1, 1:1, 1:2, 1:3, 1:5, and 1:10), ranging from very selective, to no
selectivity, creating an ``adapted'' ROC curve. Since each point is an
average of the 17 subject/image-qualities, the error bars show the standard
deviation of the spread for both A and B. We choose the 1:2 cost ratio for
further analysis because it is at the inflection point of the ``adapted'' ROC
curve, and it has the closest average (A,B) to that of semi-auto. Table II and
Fig.~\ref{Table2graph}{\bf a} shows the individual results for each
subject/image-quality for the semi-auto method and the ANRA with 1:2 cost
ratio. Fig.~\ref{Table2graph}{\bf b} shows an example of semi-auto and ANRA
points compared to the gold standard.

The results show that ANRA has a significantly higher A value of recognition
(P-value: 0.002) and a similar B value of recognition compared to the semi-auto
method.

\begin{figure}[p]
\begin{center}
$\begin{array}{c} 
\includegraphics*[width=8cm]{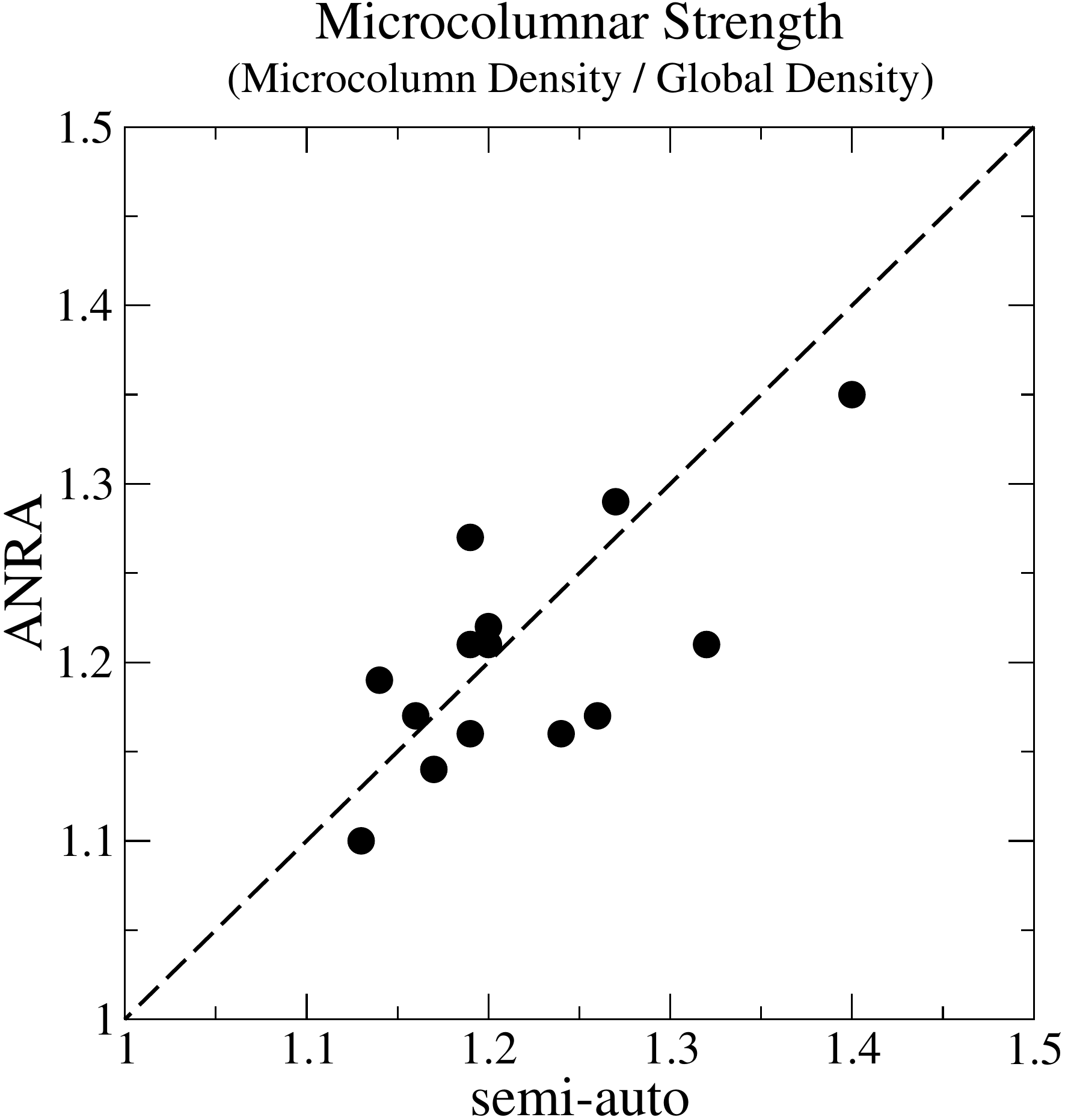}
\end{array}$
\end{center}
\caption{ Comparison of microcolumnar strength measurement ($S$) using the
$(x,y)$ locations from both ANRA (with 1:2 cost ratio) and semi-auto methods
of neruon identification. A total of 14,000 neuron locations were used, for
an average of 1000 neuron locations for each subject (plot point). Both the
neuron density and microcolumnar strength show significant correlations of
ANRA with the semi-auto method.}
\label{S.results}
\end{figure}

We also compare microcolumnar strength $S$ (Sec.~\ref{sec:dmm}) using the
$(x,y)$ locations from both ANRA and semi-auto methods of neuron
identification for the entire image database of rhesus monkey subjects as
described in Sec.~\ref{sec:methods:2}. 14,000 neuron locations were used, for
an average of 1000 neuron locations for each subject. We find significant
correlations between microcolumnar strength measurements of the ANRA and
semi-auto methods of neuron recognition (Fig.~\ref{S.results}). This shows
that ANRA has the ability to find significant changes in advanced neuron
spatial arrangements within different subjects, and can therefore be applied
to large datasets where manual or semi-auto recognition are not viable.

\bigskip 

\section{\label{sec:discuss}Discussion}

In the present work we introduce a method called an Automated
Neuron Recognition Algorithm (ANRA) which uses a combination of image
segmentation and machine learning to retrieve neuron locations within
digitized images of Nissl-stained Rhesus monkey brain tissue.  Despite
challenges, such as overlapping of neuron cell bodies and the presence of
glial cells and artifacts in the tissue, we demonstrate that ANRA has a significantly better recognition capability than a semi-auto method~\citep{Cruz:2005} which requires
expert manual intervention for each image.  ANRA's recognition quality is combined with
computational efficiency, resulting in recognition of $\sim$100 neurons per
minute using a standard personal computer. Consequently, large numbers of
neuron locations can be retrieved, spanning considerably larger brain regions
than ever before.  Furthermore, because ANRA is capable of efficiently
extracting neuron locations from durable and commonly used Nissl-stained
tissue, it can potentially be applied to vast stores of archival material
existing in laboratories and research collections around the world.

Such a large dataset of (x,y) neuron locations will allow for a variety of
systematic analyses that have previously not been possible. The ability to
identify every neuron in entire sections of the brain will allow for both
global and local analyses of neuron numbers, glial cell numbers, regional
cell densities, and local variations in cell densities. Also, as was shown in
the Results section, studies of microcolumnarity or other spatial features of
cortex, including spatial inter-relationships among neurons and glia using
autocorrelation and cross-correlation, are possible. Lastly, ANRA also allows
for less obvious applications, including the investigation of the spatial
network of the brain using the neuron locations as nodes. None of these
studies are possible with the elegant sampling methods of modern stereology.

We highlight the need for large datasets of neuron locations ($10^3-10^4$) in
comparative studies proposed in the Introduction and defined in
Sec.~\ref{sec:dmm}. Generally, the goal of a comparative study is to find a
statistically significant difference in a measured quantity (i.e.,
microcolumn strength) due to a change in an independent variable (age,
species, sex, disease state, etc.). In the case of a 1D correlation between
nearest neighbors or the 2D microcolumnar analysis, the neuron locations are
used to create 1D and 2D histograms, respectively. The number of neurons must
be high enough to resolve the effect of the independent variable above random
noise of the histogram. \citet{Buldyrev:2000} showed that for a resolution of
interest (seeing 3\% changes between 10 micron bins), $\sim10^4$ neuron
locations are needed in the comparative study of microcolumnarity. For the
same resolution in a 1D correlation comparative study, such as
nearest-neighbor distances, only {$\sim$}1000 neurons are
needed~\citep{Schmitz:2002}. For a given bin size, the theoretical
calculation shows that the required number of neurons scales as a power of
dimensions that are being correlated. Thus, automatic recognition becomes
critical in higher dimension correlations. As an example we consider a 30
subject study of neuron spatial arrangement using $\sim10^5$ neuron
locations, making 100 different measurements of 1000 neurons each through a
certain layer across several Brodmann regions.  The semi-automatic approach,
which allows for acquisition of 10 neurons per second, would take 83 human
hours to complete. Comparatively, ANRA could complete the same task in 24
hours on 20 Intel P4 processors with less than 1 hour of preparation time.

ANRA has a further advantage of reducing experimental drift. Specifically, in
terms of human bias, the ``criteria'' for neuronal identification will
necessarily differ between different observers that are often required for a
huge analysis extending over months to years, while ANRA's criteria, once
established from the training algorithm, remains constant.  Furthermore,
ANRA's criteria will not be subject to the kind of experimental drift that
can occur over time when one observer manually identifies thousands of
neurons over a period of weeks to months.

Recently, there have been advances in level set methods to recognize
overlapped cell nuclei~\citep{Lin:2007,Lockett:1999}. The recognition
challenges with Nissl-stained tissue are far greater than the challenges using
confocal microscopy using fluorescence. \citet{Lin:2007} show how neurons and
glia cells completely separate into two regions of parameter space using only
two parameters (texture and intensity) of the identified segmentations. If
plotted in a similar way, no two parameters that we consider (size,
intensity, texture, gyration, edge vs. area, etc.) would yield such a
separation. Thus, in a Nissl-stained tissue visualized by optical microscopy,
the parameterized method of \citet{Cremers:2000}, which, by design, overcomes
the challenges of noisy images and missing boundary data (Sec.~\ref{0b}), is
most efficient.

Our results suggest that the ANRA method is performing as maximal efficiency:
when a second expert's marks are compared with the gold standard on the same
Nissl-stained image, the performance ($A=88\pm5\%$) is not significantly higher than ANRA's performance ($A=86\pm5\%$).

Although there are 10 free parameters within the algorithm, only two of them
called the {\it primary parameters} must be explored to find the correct
values for proper segmentation. These {\it primary parameters} are
automatically found in the OSM parameter search during training. The other
eight free parameters, which we call the {\it secondary parameters}, can be
fixed for the general task of identifying elliptical features within noisy
images with missing boundary data, thereby solidifying them for the broadly
applicable problem of neuron recognition in all Nissl-stained tissue. For a
given morphological feature of interest, once a small set of representative
images have been trained to, the training and parameters can be reused, due
to the normalization of images of different quality. This setup will allow
for the study of large areas of montaged images, or large datasets of
hundreds of slides, all with the same training. Furthermore, the free
parameters and training can be adapted for identification of other types of
neurons, glial cells, etc.

Lastly, because of the modular nature of the method (Fig.~\ref{flow}), it
will be relatively easy to replace partial aspects of the overall algorithm
by considering advances in recent published work. For example,
\citet{Tscherepanow2:2006} independently developed a method to identify
living cells that uses a larger set of training properties that is reduced
with principle/independent component analysis, and \citet{Costa:2006} has
applied advanced pattern matching to the identification of neuron cell bodies
in Nissl-stained tissue. By replacing the respective aspects of ANRA with
such methods, the ideal overall identification algorithm can be found for not
only the recognition of neuron cell bodies, but also the recognition of other
objects of scientific interest, for example living cells or glial cells.

\section{\label{sec:software}Software}

The ANRA software is available at \url{http://physics.bu.edu/~ainglis/ANRA/}.

\section{\label{sec:akn}Acknowledgments}

This work was supported by the National Institutes of Health grants
R01-AG021133, P01-AG00001 and P51-RR00165, the Alzheimer Association, and the
Bechtel Foundation.

\newpage
\clearpage

\bibliography{./mybib.withinits.use.abbr}  
\bibliographystyle{JMIunofficial} 

\newpage
\clearpage

\bigskip 
\begin{center}

{\bf Table I}\\
\smallskip
\begin{tabular}{|c|c|c|c|}
\hline
&{\bf description} &{\bf equation }\\
\hline
$v_1$ & segment area & $A_c$ \\ 
\hline
$v_2$ & average optical density ($\overline{f})$ & $\frac{1}{A_c}\sum^A{f_{ij}}$ \\ 
\hline
$v_3$ & variance of optical density & $\frac{1}{A_c}\sum^A{\left(f_{ij}-\overline{f}\right)}^2$ \\
\hline
$v_4$ & radius of gyration of optical density & $\frac{1}{A_c}\sum^A{{\left|r_{ij} - r_{c}\right|}f_{ij}} $ \\
\hline
$v_5$ & segment edge length ($\left|C\right|$) vs. segment area & ${\left|C\right|}/{A_c}$ \\ 
\hline
$v_6$ & average gradient of segment edge & $\frac{1}{\left|C\right|}{\sum^{C}\left|{{\nabla}f_{ij}}\right|^2}$ \\
\hline
$v_7$ & average change in gradient of segment edge & $\frac{1}{\left|C\right|}\sum^{C}\left|{\nabla}f_{i+1 j}\right|^2-\left|{\nabla}f_{ij}\right|^2$ \\
\hline
\end{tabular}

\end{center}

\newpage
\clearpage

\begin{center}

{\bf Table II}\\
\smallskip
\begin{tabular}{c|c c|c c}
\hline
& semi-auto  & & ANRA  \\
{\bf \# } & A(\%) & B(\%) & A(\%) & B(\%)  \\ 
\hline
1 & 81 & 	13 & 	82 & 	11 \\ \hline
2 &  71 & 	14 & 	84 & 	7 \\\hline
3 &  82 & 	14 & 	91 & 	21 \\\hline
4 &  79 & 	15 & 	78 & 	4 \\\hline
5 &  90 & 	43 & 	92 & 	30 \\\hline
6 &  65 & 	3 & 	85 & 	16 \\\hline
7 &  76 &      18 & 	87 & 	21 \\\hline
8 &  83 &       15 & 	88 & 	6  \\\hline
9 &  76	 & 	12 & 	93 & 	15 \\\hline
10 &  79 & 	10 & 	85 & 	23 \\\hline
11 &  73 & 	6 & 	77 & 	7 \\\hline
12 &  82 & 	4 & 	88 & 	11 \\\hline
13 &  92 & 	44 & 	84 & 	6 \\\hline
14 &  90 & 	26 & 	95 & 	16 \\\hline
15 &  80 & 	20 & 	80 & 	11 \\\hline
16 &  77 & 	23 & 	91 & 	28 \\\hline
17 &  75 & 	7 & 	86 & 	17 \\\hline

{\bf avg.} & {\bf 80}$\pm$7 & {\bf 17}$\pm12$ & {\bf 86}$\pm$5* & {\bf 15}$\pm$8 \\\hline
\end{tabular}

\end{center}
\bigskip 

\newpage
\clearpage

\newpage
\clearpage

\newpage
\clearpage

\newpage
\clearpage

\newpage
\clearpage

\newpage
\clearpage

\newpage
\clearpage

\begin{figure}[p]
\begin{center}
$\begin{array}{c|c|c}
\includegraphics*[height=2cm]{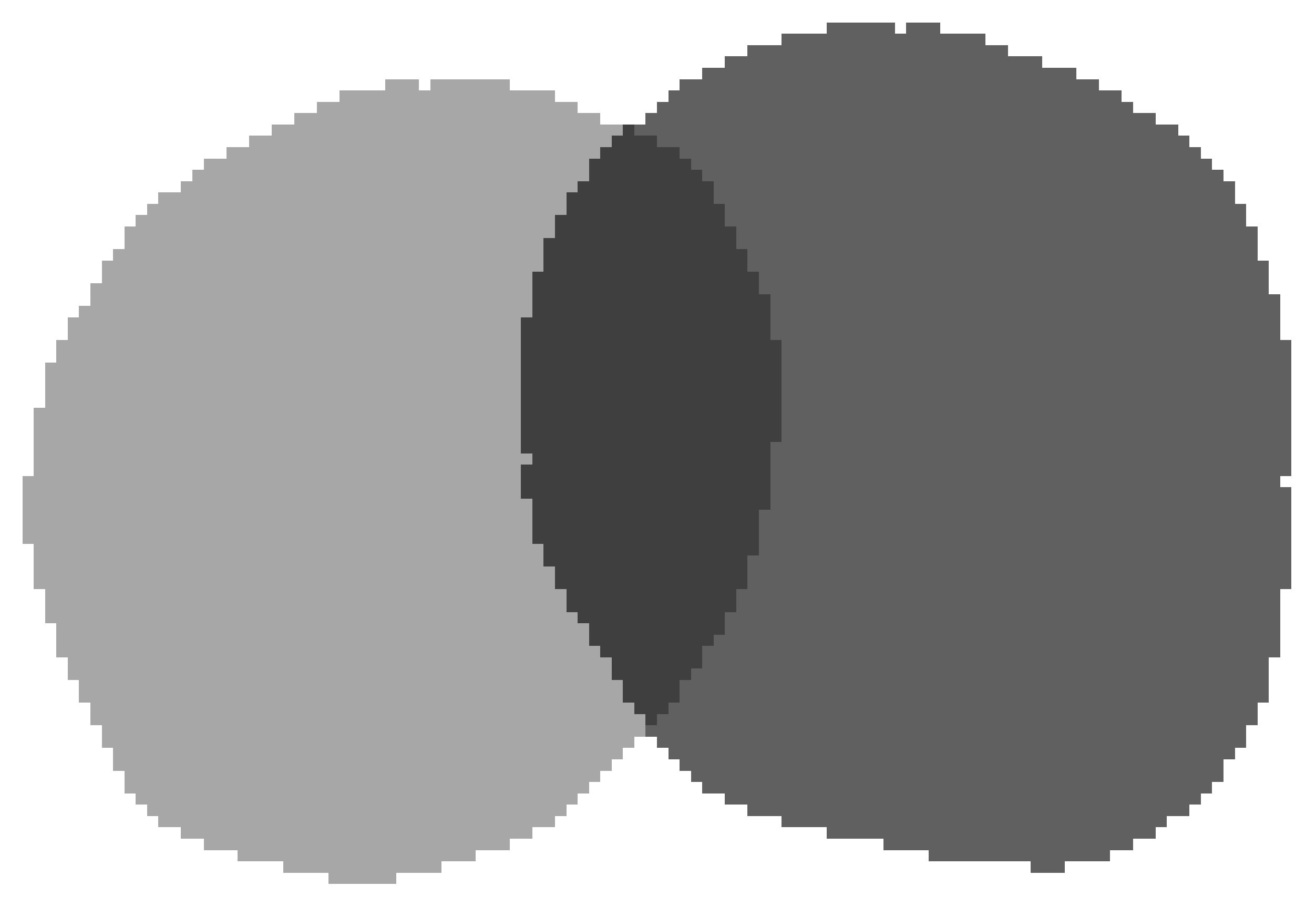} & \includegraphics*[height=2cm]{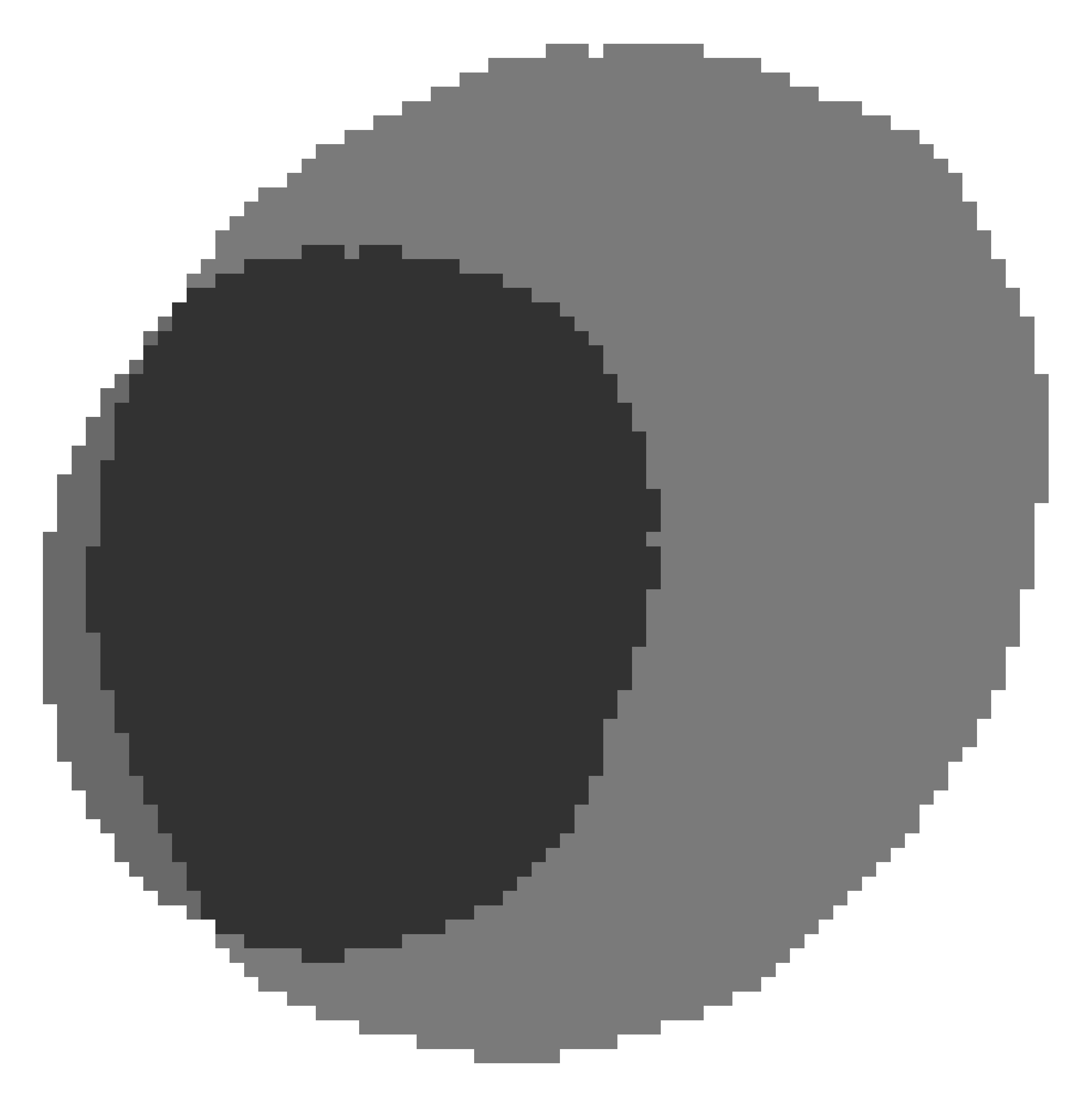} & \includegraphics*[height=2cm]{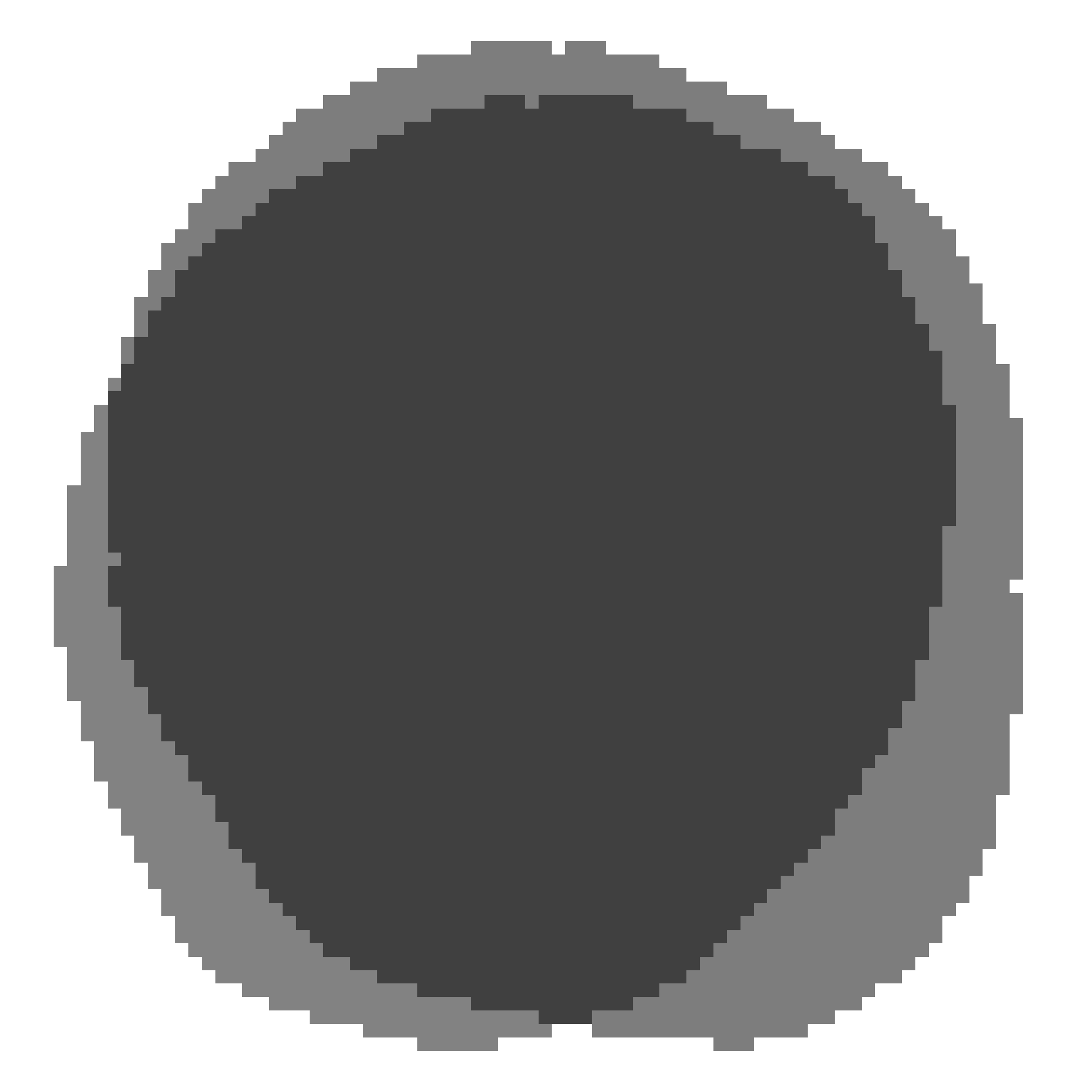} \\
\mbox{({\bf a})} & \mbox{({\bf b})} & \mbox{({\bf c})}
\end{array}$
\end{center}
\label{overlap_example}
\caption{Two segments represent the same object when they {\it mutually}
share more than 70\% of their pixels. The two segments in ({\bf a}) do not
pass the required criteria because neither segment overlaps the other by more
than 70\%. The two segments in ({\bf b}) do not pass the required criteria
because only one segment overlaps the other by more than 70\%. Only in ({\bf
c}) does the required overlap occur. This analysis is used when computer
segments are compared to ``gold standard'' training segments and either designated a
neruon or non-neuron, and during the overlap deletion phase, when the segment
with the highest probability of being a neuron is selected among all
overlapped segments.}
\label{overlap_example}
\end{figure}

\newpage
\clearpage

\newpage
\clearpage

\newpage
\clearpage

\newpage
\clearpage

\newpage
\clearpage

\end{document}